\documentclass[11pt]{article}
\usepackage[colorlinks=true,citecolor=blue,linkcolor=blue]{hyperref}
\usepackage{multirow}
\usepackage[left=2cm,right=2cm,top=2cm,bottom=2cm]{geometry}
\usepackage{cite}
\usepackage{psfrag}
\usepackage{graphicx}
\usepackage{graphics}
\usepackage{epsfig}
\usepackage{amsmath,amsfonts,amssymb}
\usepackage{pstcol,pst-fill,pst-grad}
\usepackage{euscript}
\usepackage{pstricks}
\usepackage{wrapfig}
\usepackage{slashed}
\usepackage[toc,page]{appendix}
\usepackage{here}

\date{}
\begin{document}
	\title{\vspace{-3cm}
		\hfill\parbox{4cm}{\normalsize \emph{}}\\
		\vspace{1cm}
		{ Laser-assisted kaon decay and CPT symmetry violation}}
	\vspace{2cm}
	
	\author{M Baouahi$^{1}$, M Ouali$^{1}$, M Jakha$^{1}$, S Mouslih$^{2}$, Y Attaourti$^3$, B Manaut$^1$, S Taj$^{1,}$\thanks{Corresponding author, E-mail: s.taj@usms.ma}  \\
		{\it {\small$^1$ USMS, Polydisciplinary Faculty, ERPTM, Beni Mellal, 23000, Morocco.}}\\
	{\it {\small$^2$ FST-BM, Laboratory of Materials Physics (LMP), 23000, Morocco.}}\\
	{\it {\small$^3$ High Energy Physics and Astrophysics Laboratory,
FSSM, UCAM, Morocco.}}		
	}
	\maketitle \setcounter{page}{1}
\date{\today}
\begin{abstract}
In this paper, we have investigated the charged kaons decay at the lowest order in the presence of a circularly polarized laser field. To be more precis, we have examined the leptonic decay of both positive (matter) and negative (antimatter) kaon which weakly decay via the exchange of $W$ boson. Indeed, we have derived the expression of the leptonic decay width, the leptonic branching ratio, the leptonic ratio and the charged kaon lifetime by using the decay matrix approach. In addition, by using numerical computation, we have presented and discussed how the laser field influences these physical quantities. Moreover, we have analyzed the effect of the laser field on the parameter associated to the CPT symmetry. Then, we have concluded that, in the presence of an electromagnetic field and based on this CPT symmetry parameter, it is possible to control the dominance of matter over anti-matter or vice-virsa by applying an external field to either violate or conserve the CPT symmetry.
\end{abstract}
\section{Introduction}
The great progress in laser technology since its introduction in 1960, especially by increasing its intensities and shortening its pulse duration, have received a great scientific interest\cite{Quantum_process,History,Dahiri_ElAsri:2021anl,Mouslih_and_Jakha,Ouhammou_et_al:2021,laser_prosses,laser_prosses1}. This interest is due to many previously unknown phenomena that are caused by the application of the laser radiation, allowing us to understand not only the atomic and molecular structure of matter but also phenomena related to high energy processes and particles' behaviors\cite{Mouslih_and_Jakha,Ouhammou_et_al:2021}. Moreover, laser field can pave a new ways to high energy physics. Therefore, the study of various aspects of the influence of the electromagnetic radiation on high energy scattering and decay processes is one of the most important issues of modern physics.

The study of quantum effects in laser-matter interactions may allow us to obtain new information about the nature of the interactions of the particles involved in the scattering and decay processes. Moreover, the dependence of these processes on laser field's parameters that can be easily controlled in experiments may lead to the detection of new effects that do not exist in the case of laser-free interactions between particles. In some processes, the electromagnetic field is a determining element and without it they do not occur (see for example \cite{Quantum_process,laser_prosses}), while in other processes that occur in the absence of the laser field, they are modified by its presence \cite{History,Dahiri_ElAsri:2021anl,Mouslih_and_Jakha,Ouhammou_et_al:2021}.

The precise study of rare processes where kaons play a predominant role since their discovery in 1947 \cite{discovery_of_kaons} gives us today the light to understand the structure of the weak interaction. In particle physics, CP symmetry violation has been one of the most sensitive issues since its discovery in 1964 in neutral kaon decays \cite{Christenson:1964fg} and later in B mesons\cite{CP_viol_in_B_2001,CP_viol_in_B_2004}. Therefore, it is expected that the weak interaction violating CP symmetry also violates the invariance of time inversion T which was found in $B$ mesons when exchanging initial and final states in transitions which can be connected by a T symmetry transformation \cite{Lees:2012uka}. In this context, an elegant explanation of this effect was proposed by Kobayashi and Maskawa, as a CP violation phase in the Cabibbo-Kobayashi-Maskawa (CKM) three-generation quark mixing matrix\cite{CKM:1963_1973} in the Standard Model (SM). So with this phenomena related to the CP symmetry characterizing this interaction \cite{Christenson:1964fg}, the question arises as to whether there exists an effect capable of controlling unique properties of an interaction (weak interaction) by another interaction (electromagnetic interaction), and what are the tangible variations of the measurable quantities characterizing the interacting particles. For that, the process of the charged kaons decay into two leptonic bodies, which is recognized in the Standard Model by the exchange of a bosonic propagator $W$ to give two leptons, is developed in this work. The laser-assisted decay process is the preferred choice for charged particles not only because of the fact that the electromagnetic field interacts only with charged particles but also for that this decay process may give us the possibility to extend the particles' lifetime. We note that at high laser field's intensites where the Schwinger limit is respected $\epsilon_{0}=(m^{2}c^{3})/(e\hbar)=1.32 \times 10^{16} V/cm$ \cite{Sauter:1931,Schwinger:1951}, the two leptonic branching ratios are affected in such a way that they will be of the same order.

CP violation allows the two partial decay widths of the particle and antiparticle to be different even if CPT symmetry is valid. In addition, CPT symmetry also imposes equality of lifetimes for CP-conjugate states. In the case where the produced particles interact only via weak interaction or if the final state interactions does not involve transitions between different decay channels, CPT invariance is sufficient to ensure equality of the partial rates. Therefore, the comparison of leptonic decay widths of the charged kaon is a test of CPT invariance \cite{CPT,Sozzi:2008,Ford:1967zz}. For this reason, we will use the asymmetry parameter defined in the theoretical part.

In this work, the laser-assisted charged kaon decay process is studied without taking into account the radiative corrections and by considering the kaon as a structureless particle. Moreover, the external field, which is considered to be monochromatic and circularly polarized, will influence all charged particles of the initial and final states and exchanges multiphotons \cite{Ritus:1985}. This effect of the laser field is examined through the theoretical measurement of measurable quantities such as that associated with the CPT symmetry, the partial decay width, the leptonic ratio, the leptonic branching ratio and the lifetime of the charged kaons. We consider, for the kaon decay process without laser field, that matter and antimatter have the same mass, and also their found decay widths are equal. This equality ensures the conservation of CPT symmetry at the lowest order.

The remainder of this research paper is organized as follows: The section (\ref{Sec.1}) is about the theoretical analysis of some quantities which characterize the two body leptonic charged kaons decay process at the lowest order in the absence and in the presence of a circularly polarized electromagnetic field. The obtained results and data are presented and analyzed in section (\ref{Sec2}). In section (\ref{Sec3}), we have given a short conclusion. Some calculation's details of the charged kaon decay width are given in the appendix.  We mention that we have used natural units such that $\hbar=c=1$, and the signature metric is taken as $(+---)$.
\section{Outline of theory}\label{Sec.1}
This paper's part is devoted to the analytical calculation of the charged kaon partial decay width as it is one of the most important measured quantities in decay processes. It is well known that the charged kaon may decays via several channels \cite{PDG:2020}. However, these decays modes are different in terms of their decay rates. We will focus on the leptonic decay modes at the lowest order which are $K^{+}\rightarrow l^{+}+\nu_{l}$ and $K^{-}\rightarrow l^{-}+\overline{\nu}_{l}$. Indeed, we begin by giving the expression of the laser-free decay width then we will move on to derive its expression in the presence of an electromagnetic field.
\subsection{Laser-free kaon decay}
In general, the leptonic tow-body decay width of a charged mesons $ P^{c} $, such as $ \pi^{\pm} $, $ K^{\pm} $, $ Ds^{\pm} $ and $ B^{\pm}$ , in the absence of the electromagnetic field is given by:
\begin{eqnarray}\label{decay_without_laser}
\Gamma &=&\dfrac{G_{F}^{2}F_{P^{c}}^{2}}{4\pi}m_{1}m_{l}^{2}(1-\dfrac{m_{l}^{2}}{m_{1}^{2}})^{2},
\end{eqnarray}
where $G_{F}$ and $F_{P^{c}}$ are successively the Fermi constant and the charged meson's structure constant. $ m_{1} $ is the mass of the charged mesons $P^{c}$. For the charged kaon decay, we have $ m_{1}=m_{K^{\pm}}=493.677\,MeV$. We notice that the mass of neutrinos is taken to be zero. $m_{l}$ is the mass of the produced charged lepton such that: $ m_{l}=m_{\mu^{\pm}}=105,6583745\,MeV$ for the muon and the antimuon, or $ m_{l}=m_{e^{\pm}}=0.5109989\,MeV$ for the electron and the positron.
\subsection{Laser-assisted kaon decay}
The leptonic decay of the charged kaon in the presence of an external electromagnetic field can be described as follows:
\begin{eqnarray}
K^{\pm}(q_{1})\rightarrow l^{\pm}(q_{2})+\nu_{l_{i}}(P_{3}),
\end{eqnarray}
where $l$ can be either an electron or a muon, and $\nu_{l_{i}}$ represents $\nu_{l}$ for $K^{+}$ decay and $\bar{\nu_{l}}$ for $K^{-}$ decay. We consider that the charged kaon is initially represented by the 4-vector momentum $P_{1}=(E_{1},\vec{0})$, but when it is dressed by the laser field, its 4-vector momentum becomes the effective 4-vector momentum $ q_{1}=P_{1}-(e^{2}a^{2})/(2k.P_{1})k $, where $ k=(\omega,\vec{k}) $, $ e $ and $ a $ are respectively the electromagnetic wave 4-vector, the charge of the kaon and the polarization of the laser field. $ q_{2}=P_{2}-(e^{2}a^{2})/(2k.P_{2})k $ is the effective 4-vector momentum of the lepton $ l^{\pm} $, and $P_{3}=(E_{3},\vec{P_{3}})$ is the 4-vector momentum of $ \nu_{l_{i}} $. The classical four-vector potential which represents the circularly polarized monochromatic laser field is given by:
\begin{eqnarray}
 A^{\mu}(\phi)=a_{1}^{\mu}\cos\phi+a_{2}^{\mu}\sin\phi,
\end{eqnarray}
with $ \phi=(k.x) $ is the phase of the laser field. The 4-vectors $ a_{1} $ and $ a_{2} $ are orthogonal, and chosen such that $ a_{1}^{\mu}=|a|(0,1,0,0) $ and $ a_{2}^{\mu}=|a|(0,0,1,0) $. Thus, they verify the following conditions: $a_{1}. a_{2}=a_{2}. a_{1}=0$ and $ a_{1}^{2}=a_{2}^{2}=a^{2}=-|a|^{2}=-(\epsilon_{0}/\omega)^{2} $, with $ \epsilon_{0} $ is the amplitude of the laser's electric field.
The produced charged lepton inside the laser field is described by the analytical solution of Dirac equation (Dirac-Volkov wave function)\cite{Volkov:1935} such that:
\begin{eqnarray}
\psi_{l^{\pm}}(x)=\left[ 1+\dfrac{e\slashed{k}\slashed{A}}{2 k.P_{2}}\right] \dfrac{DS(P_{2},s_{2})}{\sqrt{2Q_{2}V}}e^{\mp iS(q_{2},x)},
\end{eqnarray}
$ V $ is the quantization volume. In general, $ DS(P_{i},s_{i}) $ represents the Dirac bispinor  which can be either $ u(P_{i},s_{i}) $ for lepton $ l^{-} $ or $ v(P_{i},s_{i}) $ for antilepton $ l^{+} $. It verifies the following relations:
\begin{eqnarray}\label{Dirac_spinor}
\sum_{s_{i}}u(P_{i},s_{i})\bar{u}(P_{i},s_{i})=\slashed{P_{i}}+m_{l} \qquad\text{and}\qquad \sum_{s_{i}}v(P_{i},s_{i})\bar{v}(P_{i},s_{i})=\slashed{P_{i}}-m_{l},
\end{eqnarray}
where $ s_{i} $ and $ m_{l} $ denotes successively the particle's spin and its mass.
The classical action of the produced charged lepton in the plane electromagnetic wave (phase of the Dirac-Volkov state) is given by the following expression:
\begin{eqnarray}
S(q_{2},x)=q_{2}.x-e\dfrac{a_{1}.P_{2}}{k.P_{2}}\sin\phi+e\dfrac{a_{2}.P_{2}}{k.P_{2}}\cos\phi,
\end{eqnarray}
The square of the effective four-vector momentum $ q_{2}=(Q_{2},\vec{q_{2}}) $, where its zero component $ Q_{2}$ is the charged lepton's effective energy, is defined such that:
\begin{eqnarray}
{q_{2}}^{2}={P_{2}}^{2}-e^{2}a^{2}=m_{l}^{2}-e^{2}a^{2}= {m_{l}^{*}}^{2},
\end{eqnarray}
where $ m_{l}^{*} $ is the charged lepton's effective mass.
Inside the laser field, the charged kaon is described by the Klein-Gordon equation of a charged scalar coupled to a classical electromagnetic field\cite{Volkov:1935,Berestetskii_et_al:1982}, which is given by the following expression.
\begin{eqnarray}\label{K-G-WL}
\left[ (i\partial-eA)^{2}-m_{K}\right]\psi_{K}(x)=0,
\end{eqnarray}
where $ m_{K} $ is the charged kaon mass. From the equation (\ref{K-G-WL}), we can derive the wave function of the charged kaon as follows:
\begin{eqnarray}
\psi_{K^{\pm}}(x)&=&\dfrac{1}{\sqrt{2Q_{1}V}}e^{\pm iS(q_{1},x)},
\end{eqnarray}
where the classical action  $S(q_{1},x) $ is expressed as follows:
\begin{eqnarray}
S(q_{1},x)=q_{1}.x-e\dfrac{a_{1}.P_{1}}{k.P_{1}}\sin\phi+e\dfrac{a_{2}.P_{1}}{k.P_{1}}\cos\phi.
\end{eqnarray}
The square of the effective four-vector momentum $ q_{1}=(Q_{1},\vec{q_{1}}) $ is given by:
\begin{eqnarray}
{q_{1}}^{2}={P_{1}}^{2}-e^{2}a^{2}=m_{K}^{2}-e^{2}a^{2}= {m_{K}^{*}}^{2}.
\end{eqnarray}
We mention that the neutrino / anti neutrino is considered as massless particle. It is a neutral particles which means that it doesn't interact with the electromagnetic field. Indeed, it is described by a free wave function as follows:
\begin{eqnarray}
\psi_{\nu_{l_{i}}}(x)&=&\dfrac{DS(P_{3},s_{3})}{\sqrt{2E_{3}V}}e^{\mp iP_{3}.x},
\end{eqnarray}
where $ E_{3} $ represents the neutrino's energy.
In the first Born approximation and by following Feynman's rules, the matrix elements  $S_{fi} $ of the leptonic decay modes of the charged kaon ($S_{fi}^{+} $ for the matter decay and $S_{fi}^{-} $ for the antimatter decay) can be expressed as follows \cite{Greiner_Muller:1982}:
\begin{eqnarray}\label{Transition_Element}
S_{fi}^{\pm}=\dfrac{-iG_{F}}{\sqrt{2}}\int d^{4}x\mathcal{J}_{\mu}^{K^{\pm}\dagger}(x)\mathcal{J}^{l^{\pm} \mu}(x),
\end{eqnarray}
where $ G_{F}=(1.16637\pm0.00002)\times 10^{-11}\,MeV^{-2}$ is the Fermi coupling constant. The laser-assisted leptonic currents for matter decay ($\mathcal{J}^{l^{+}}$) and antimatter decay ($\mathcal{J}^{l^{-}}$) are given by:
\begin{eqnarray}\label{Leptonic_current}
\mathcal{J}^{l^{+} \mu}(x)=\bar{\psi}_{l^{+}}(x)\gamma^{\mu}(1-\gamma^{5})\psi_{\nu_{l}}(x) \qquad\text{and}\qquad\mathcal{J}^{l^{-} \mu}(x)=\bar{\psi}_{l^{-}}(x)\gamma^{\mu}(1-\gamma^{5})\psi_{\bar{\nu_{l}}}(x).
\end{eqnarray}
The associated laser-assisted hadronic currents are expressed as follows :
\begin{eqnarray}\label{Hadronic_current}
\mathcal{J}_{\mu}^{K^{\pm}}(x)=i\sqrt{2}F_{K}P_{1\mu}\dfrac{1}{\sqrt{2Q_{1}V}}e^{\pm iS(q_{1},x)},
\end{eqnarray}
where the kaon decay constant $ F_{K}$ is obtained from the leptonic decay modes of the charged kaon in the absence of the laser field \cite{Greiner_Muller:1982} such that: $ F_{K}=24.9414\,MeV$ for the muonic channel and $ F_{K}=24.5501\,MeV$ for the electronic channel. After calculation and by substituting the equations (\ref{Hadronic_current}) and (\ref{Leptonic_current}) into equation (\ref{Transition_Element}), the decay matrix element of the two leptonic decays become as follows:
\begin{eqnarray}\label{smatrix1}
S_{fi}^{\pm}&=&\dfrac{-G_{F}F_{K}}{\sqrt{8Q_{1}Q_{2}E_{3}V^{3}}}\int d^{4}x\bar{DS}(P_{2},s_{2})(1+c(P_{2})\slashed{a_{1}}\slashed{k}\cos\phi+c(P_{2})\slashed{a_{2}}\slashed{k}\sin\phi) \nonumber \\
&\times &\slashed{P_{1}}(1-\gamma^{5})DS(P_{3},s_{3})e^{i(\mp S(q_{1},x)\pm S(q_{2},x)\mp P_{3}.x)},
\end{eqnarray}
with $c(P_{2})=\dfrac{e}{2k.P_{2}}$. The exponential term in equation (\ref{smatrix1}) can be evaluated as follows:
\begin{eqnarray}
\pm\left[ -S(q_{1},x)+S(q_{2},x)-P_{3}.x\right] =\pm\left[ (q_{1}-q_{2}-P_{3}).x+z\sin(\phi-\phi_{0})\right],
\end{eqnarray}
where $z$ and $\phi_{0}$ are defined as:
\begin{eqnarray}
 z=\sqrt{\alpha_{1}^{2}+\alpha_{2}^{2}}, \qquad  \phi_{0}=\arctan(\dfrac{\alpha_{2}}{\alpha_{1}}),
\end{eqnarray}
with:
\begin{eqnarray}
\alpha_{1}=e(\dfrac{a_{1}.P_{1}}{k.P_{1}}-\dfrac{a_{1}.P_{2}}{k.P_{2}}), \qquad \alpha_{2}=e(\dfrac{a_{2}.P_{1}}{k.P_{1}}-\dfrac{a_{2}.P_{2}}{k.P_{2}}).
\end{eqnarray}
By using the generating function of Bessel functions, $ e^{iz\sin\theta}=\sum_{n=-\infty}^{n=+\infty}J_{n}(z)e^{in\theta} $, we can express the charged kaon decay matrix element  $S_{fi}^{\pm} $ in terms of $Bn^{\pm}(z)$, $B1n^{\pm}(z)$ and $B2n^{\pm}(z)$ which are defined as follows:
\begin{eqnarray}\label{Bessel_transformations}
\boldsymbol{B}n^{\pm}(z)&=&J_{n}(z)e^{\mp in\phi_{0}}, \nonumber \\
\boldsymbol{B}1n^{\pm}(z)&=&\frac{J_{n+1}(z)e^{\mp i\phi_{0}(n+1)}+J_{n-1}(z)e^{\mp i\phi_{0}(n-1)}}{2}, \\
\boldsymbol{B}2n^{\pm}(z)&=&\frac{J_{n+1}(z)e^{\mp i\phi_{0}(n+1)}-J_{n-1}(z)e^{\mp i\phi_{0}(n-1)}}{2i},\nonumber
\end{eqnarray}
where $n$ is the number of exchanged photons between the electromagnetic field and the particles involved in the decay process. By using some of the Bessel function properties, we get the following transformations:
\begin{eqnarray}
 e^{\pm iz\sin(\phi-\phi_{0})}&=&\sum_{n=-\infty}^{n=+\infty}\boldsymbol{B}n^{\pm}(z),\\
\cos\phi e^{\pm iz\sin(\phi-\phi_{0})}&=&\sum_{n=-\infty}^{n=+\infty}\boldsymbol{B}1n^{\pm}(z), \\
\sin\phi e^{\pm iz\sin(\phi-\phi_{0})}&=&\sum_{n=-\infty}^{n=+\infty}\boldsymbol{B}2n^{\pm}(z).
\end{eqnarray}
By inserting these transformations in equation (\ref{smatrix1}), the decay matrix element S$_{fi}^{\pm} $ becomes as follows:
\begin{eqnarray}
{S_{fi}}^{\pm}=\dfrac{-G_{F}F_{K}}{\sqrt{8Q_{1}Q_{2}E_{3}V^{3}}}\sum_{n=-\infty}^{n=+\infty}(2\pi)^{4}\delta^{4}(\pm q_{1}\mp q_{2}\mp P_{3}\pm nk){{{\mathcal{M}}_{fi}}^{n}}^{\pm},
\end{eqnarray}
where $ {{{\mathcal{M}}_{fi}}^{n}}^{\pm} $ is the decay amplitude which is given by:
\begin{eqnarray}\label{Mfi}
{{\mathcal{M}}_{fi}}^{n\pm}&=&\bar{DS}(P_{2},s_{2})\left[ \boldsymbol{B}n^{\pm}(z)+c(P_{2})\slashed{a_{1}}\slashed{k}\boldsymbol{B}1n^{\pm}(z)+c(P_{2})\slashed{a_{2}}\slashed{k}\boldsymbol{B}2n^{\pm}(z)\right] \nonumber\\
&\times&\slashed{P_{1}}(1-\gamma^{5})DS(P_{3},s_{3}).
\end{eqnarray}
To express the partial decay width of matter ($ \Gamma^{+} $) and antimatter ($ \Gamma^{-} $), the decay matrix element is weighted by the phase space and time unit $ T  s$uch that:
\begin{eqnarray}
\Gamma^{\pm} &=&\dfrac{1}{T}\int \dfrac{Vd^{3}\vec{q}_{2}}{(2\pi)^{3}}\int \dfrac{Vd^{3}\vec{P}_{3}}{(2\pi)^{3}}\vert \bar{S}_{fi}^{\pm}\vert^{2}=\sum_{n=-\infty}^{n=+\infty}\Gamma^{n\pm},
\end{eqnarray}
where $ \Gamma^{n\pm} $ has the following expression:
\begin{eqnarray}\label{partial_decay}
\Gamma^{n\pm}=\dfrac{G_{F}^{2}F_{K}^{2}}{32\pi^{2}Q_{1}}\int d\Omega_{q_{2}}\int_{0}^{+\infty} \dfrac{\vec{q}_{2}^{2}d\vert\vec{q}_{2}\vert}{Q_{2}E_{3}}\delta(\pm Q_{1}\mp Q_{2}\mp E_{3}\pm nw)\displaystyle{\sum_{s_{2},s_{3}}}\vert \mathcal{M}_{fi}^{n\pm}\vert^{2},
\end{eqnarray}
with $ \Omega_{q_{2}} $ is the solid angle associated to the dressed charged lepton $ l^{\pm} $. The evaluation of both the integral over $ \vert\vec{q}_{2}\vert $ and the term $ \displaystyle{\sum_{s_{2},s_{3}}}\vert \mathcal{M}_{fi}^{n\pm}\vert^{2} $ in equation (\ref{partial_decay}) are given in the appendix. \\
Let's define the total decay width $\Gamma^{\pm}_{T}$ of the kaon decay at the lowest order such that:
\begin{eqnarray}
 \Gamma_{T}^{\pm} =\Gamma^{\pm}\left[ K^{\pm}\rightarrow e^{\pm} + \nu_{e_{i}}\right]+\Gamma^{\pm}\left[K^{\pm}\rightarrow \mu^{\pm}+\nu_{\mu_{i}}\right]+\Gamma_{O.C.}^{\pm},
\end{eqnarray}
where $ \Gamma_{O.C.}^{\pm} $ is the summation of the non leptonic decay widths of the charged kaons. Thus, the leptonic decay branching ratio $ BR_{l}^{\pm} $ can be expressed as follows:
\begin{eqnarray}\label{leptonic_BR}
BR_{l}^{\pm}\left[ K^{\pm}\rightarrow l^{\pm}+\nu_{l_{i}} \right] =\dfrac{\Gamma^{\pm}\left[ K^{\pm}\rightarrow l^{\pm}+\nu_{l_{i}} \right] }{\Gamma_{T}^{\pm}}.
\end{eqnarray}
Eventhough the charged kaon has fifty disintegration modes \cite{PDG:2020} and for the fact that we only have taken into account its leptonic decay modes, we shall define the lifetime $ \tau(K^{\pm}) $ which is associated to each decay process such that:
\begin{eqnarray}\label{lifetime}
\tau(K^{\pm})=\dfrac{1}{\Gamma^{\pm}_{T}}.
\end{eqnarray}
We have also defined the leptonic ratio $ R_{e/\mu} $, which is given experimentally by $ R_{e/\mu}=(2.488\pm 0.009)\times10^{-5} $, in the presence of a circularly polarized laser field as follows:
\begin{eqnarray}\label{leptonic_report}
R_{e/\mu}=\dfrac{\Gamma^{\pm}\left[ K^{\pm}\rightarrow e^{\pm}+\nu_{e_{i}}\right]}{\Gamma^{\pm}\left[ K^{\pm}\rightarrow \mu^{\pm}+\nu_{\mu_{i}}\right]}.
\end{eqnarray}
To discuss the effect of the laser field on the matter-antimatter decay, we have defined the decay factor $ \vartriangle_{CPT}(l)$, which is experimentally equal to $(-0.27\pm0.21)\%$ without laser field in the muonic channel\cite{PDG:2020}, by the following relation:
\begin{eqnarray}\label{CPT_parameter}
\triangle_{CPT}(l)=\dfrac{\Gamma^{+}\left[ K^{+}\rightarrow l^{+}+\nu_{l}\right]-\Gamma^{-}\left[ K^{-}\rightarrow l^{-}+\bar{\nu_{l}}\right]}{\Gamma^{+}\left[ K^{+}\rightarrow l^{+}+\nu_{l}\right]+\Gamma^{-}\left[ K^{-}\rightarrow l^{-}+\bar{\nu_{l}}\right]}.
\end{eqnarray}
\section{Results and discussion}\label{Sec2}
After dealing with the theoretical calculation of the laser-assisted two-body leptonic decay of the charged  mesons $  K^{\pm} $, the results and discussion part will be devoted to study the variation of different quantities and observables already illustrated in the previous section in order to understand the behavior of the charged kaon inside the laser field. We have used the FormCalc-9.3.1 \cite{feyncalc1,feyncalc2,feyncalc3} package to evaluate the spinorial part of the decay width. The different numerical results are obtained in the natural units system by using unit conversions such as: The amplitude of the electric field $ \epsilon_{0}\,[eV^{2}]=43290.844\times\epsilon_{0}\,[V/cm]$, the time $  t\,[s]=t/6.582122\times10^{-16}\,[eV^{-1}]$ and the electric charge $ |e|=8.5424546\times10^{-2} $. Throughout this work, we have chosen spherical coordinates such that the spherical angle $ \varphi $ is chosen as equal to zero, and the wave vector $ \vec{k} $ is taken as along the $ z $ axis. We begin our discussion by analyzing the variation of the differential partial decay width as a function of the number of emitted or absorbed photons.

\begin{figure}[H]
\centering
  \begin{minipage}[t]{0.45\textwidth}
  \centering
    \includegraphics[width=\textwidth]{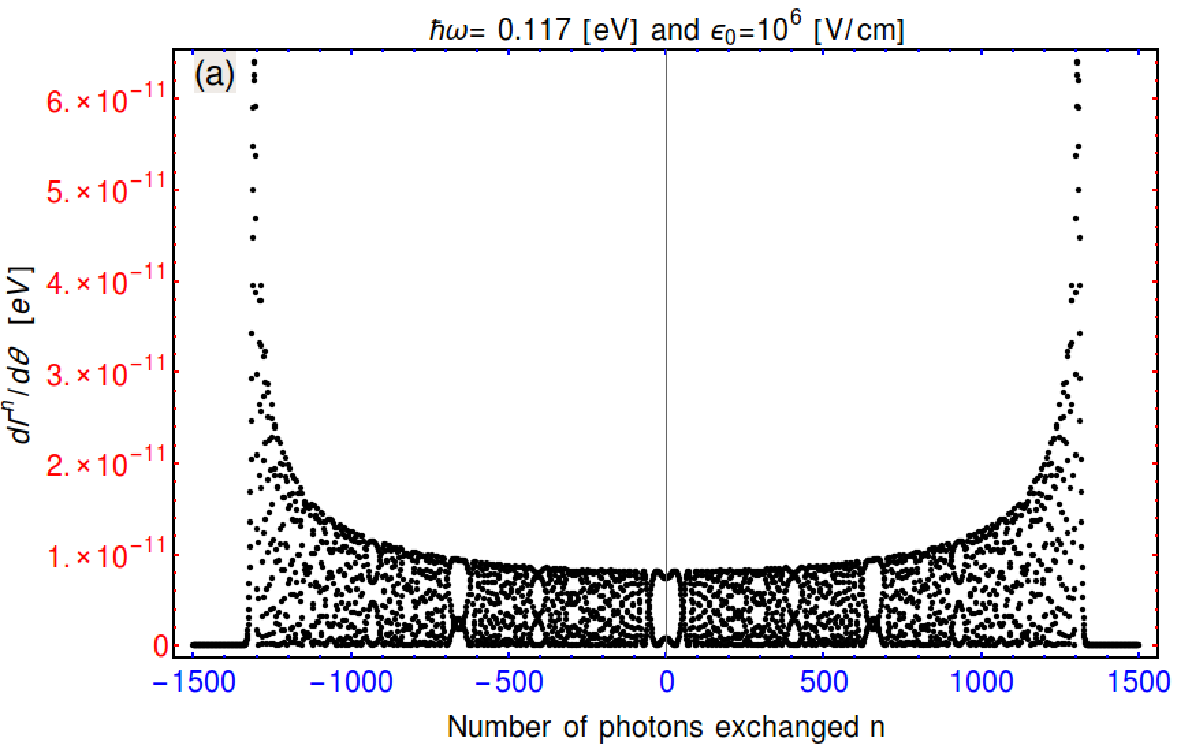}
  \end{minipage}
  \hspace*{0.25cm}
  \begin{minipage}[t]{0.45\textwidth}
  \centering
    \includegraphics[width=\textwidth]{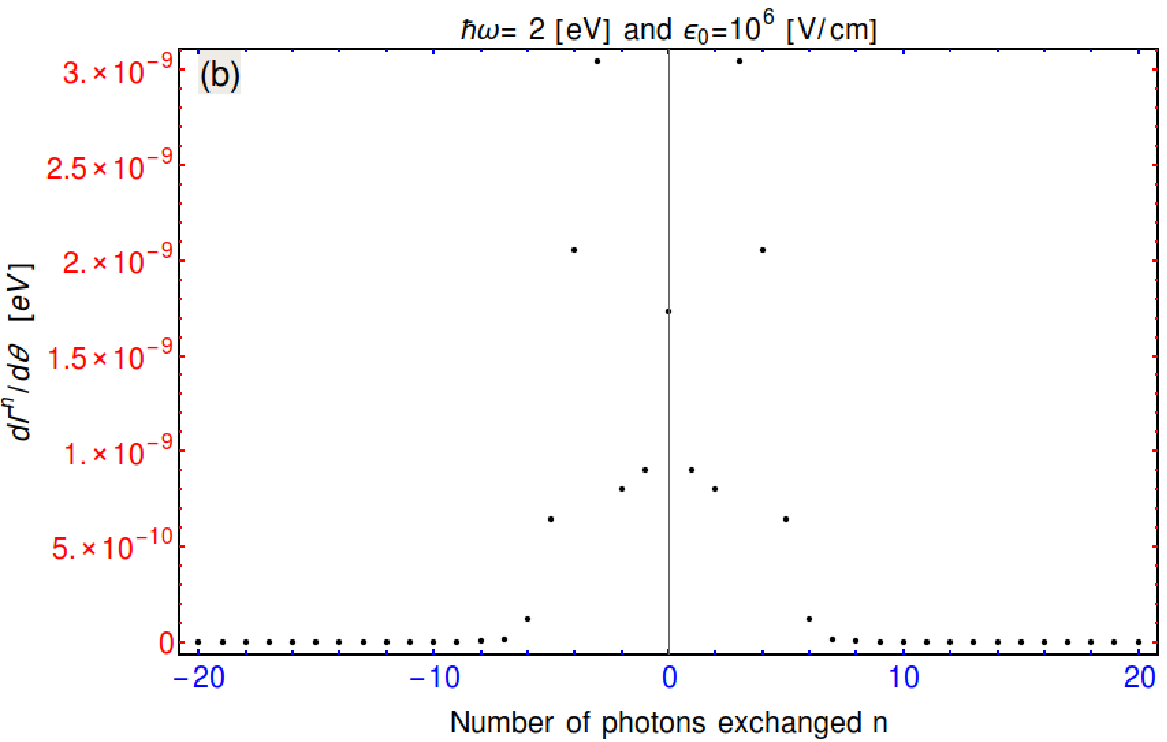}
  \end{minipage}
   \begin{minipage}[t]{0.45\textwidth}
  \centering
    \includegraphics[width=\textwidth]{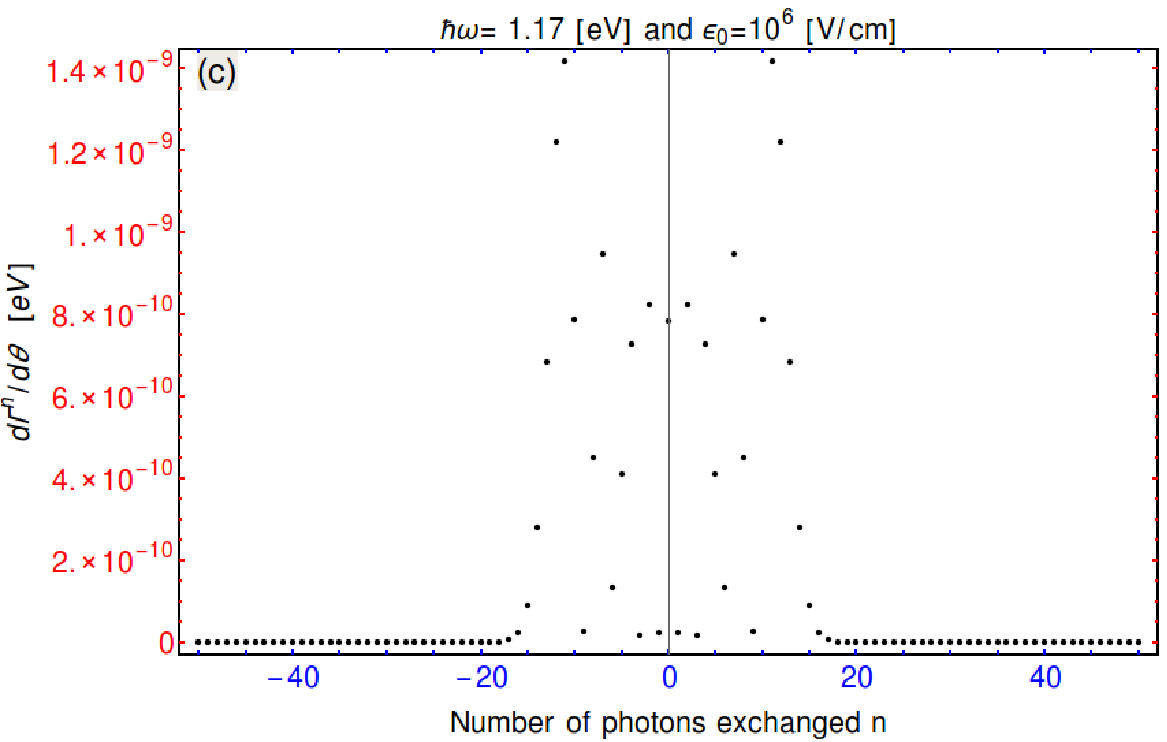}
  \end{minipage}
  \hspace*{0.25cm}
  \begin{minipage}[t]{0.45\textwidth}
  \centering
    \includegraphics[width=\textwidth]{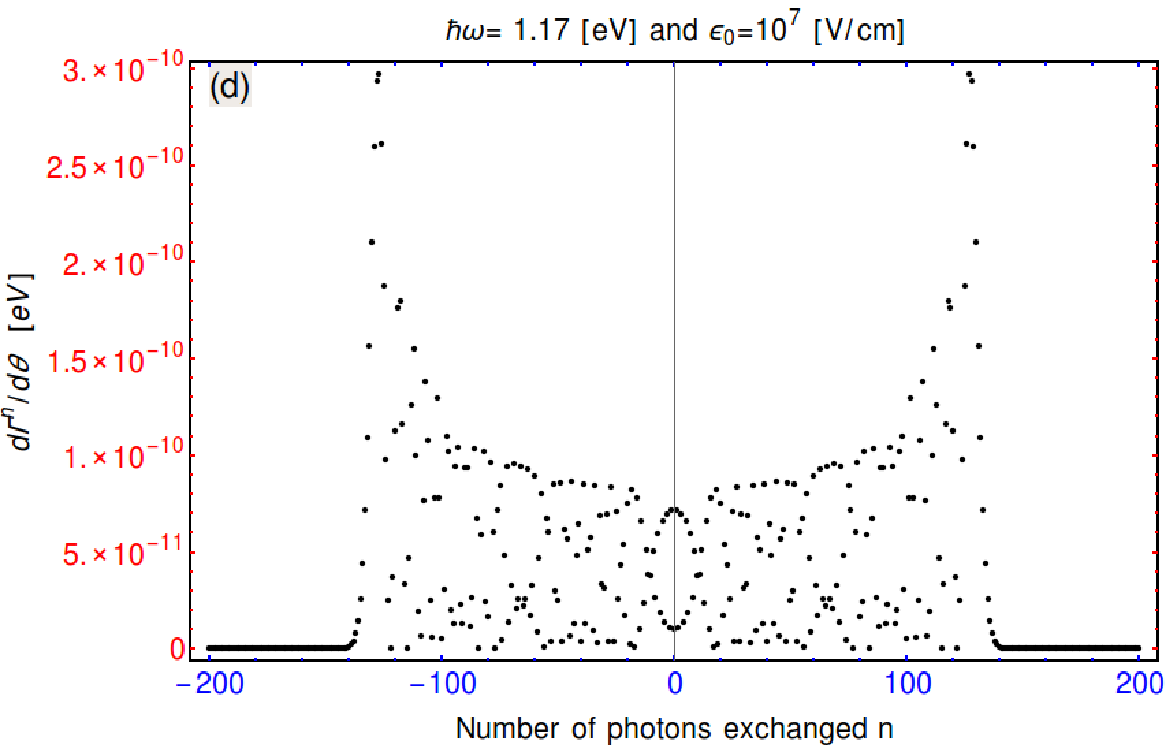}
  \end{minipage}
  \caption{Envelope of the differential partial decay width of the charged kaon $ \Gamma^{s+}( K^{+} \rightarrow\mu^{+}+\nu_{\mu}) $(\ref{partial_decay}) as a function of the photons number $n$ for different values of $\hbar\omega$ and $\epsilon_{0}$, by choosing the spherical coordinates such as $ \theta=90^{\circ} $ and $ \varphi=0^{\circ} $.} \label{Figure:1}
\end{figure}
Figure \ref{Figure:1} illustrates the dependence of the differential partial decay width of the charged kaon on the photons number $n$ for different laser field amplitudes and frequencies. This figure aims to perform two comparisons. Firstly, by comparing the two figures \ref{Figure:1}(a) and \ref{Figure:1}(b), for the same laser field strength $\epsilon_{0}= 10^{6}\,V/cm$, the laser-assisted differential multiphoton decay rate \textbf{\textbf{d$\Gamma^{n}$/d$\theta$}} depends on the laser frequency as it moves from the order of $ 10^{-11}\,eV $ for $\hbar\omega=0.117\,eV$ to $ 10^{-9}\,eV$ for the case where $\hbar\omega=2\,eV$. In addition, the number of photons that can be exchanged between the particles involved in the decay process and the laser field decreases by increasing the laser frequency. For instance, this number is approximately $n=1330$ in the case of $\hbar\omega=0.117\,eV$, while $n=9$ for $\hbar\omega=2\,eV$. Secondly, in the two figures \ref{Figure:1}(c) and \ref{Figure:1}(d), where the laser frequency is fixed at $\hbar\omega=1.17\,eV$, the differential partial decay width presents symmetric cutoffs with respect to $n=0$. Moreover, the partial decay width decreases by decreasing the laser field strength. Consequently, the cutoff number increases by either decreasing the laser frequency or increasing the laser field strength. This first result is in full agreement with that found in other studies\cite{Mouslih_and_Jakha} as it gives us information about the number of photons $n$ that can be exchanged when the laser-assisted decay width is different from its corresponding laser-free decay width. We will now focus our attention on discussing the variation of the laser-assisted differential decay width d$\Gamma^{\pm}$/d$\theta$ as a function of the angle $ \theta $ for the two leptonic modes of the charged kaon decay.

\begin{figure}[H]
\centering
  \begin{minipage}[t]{0.45\textwidth}
  \centering
    \includegraphics[width=\textwidth]{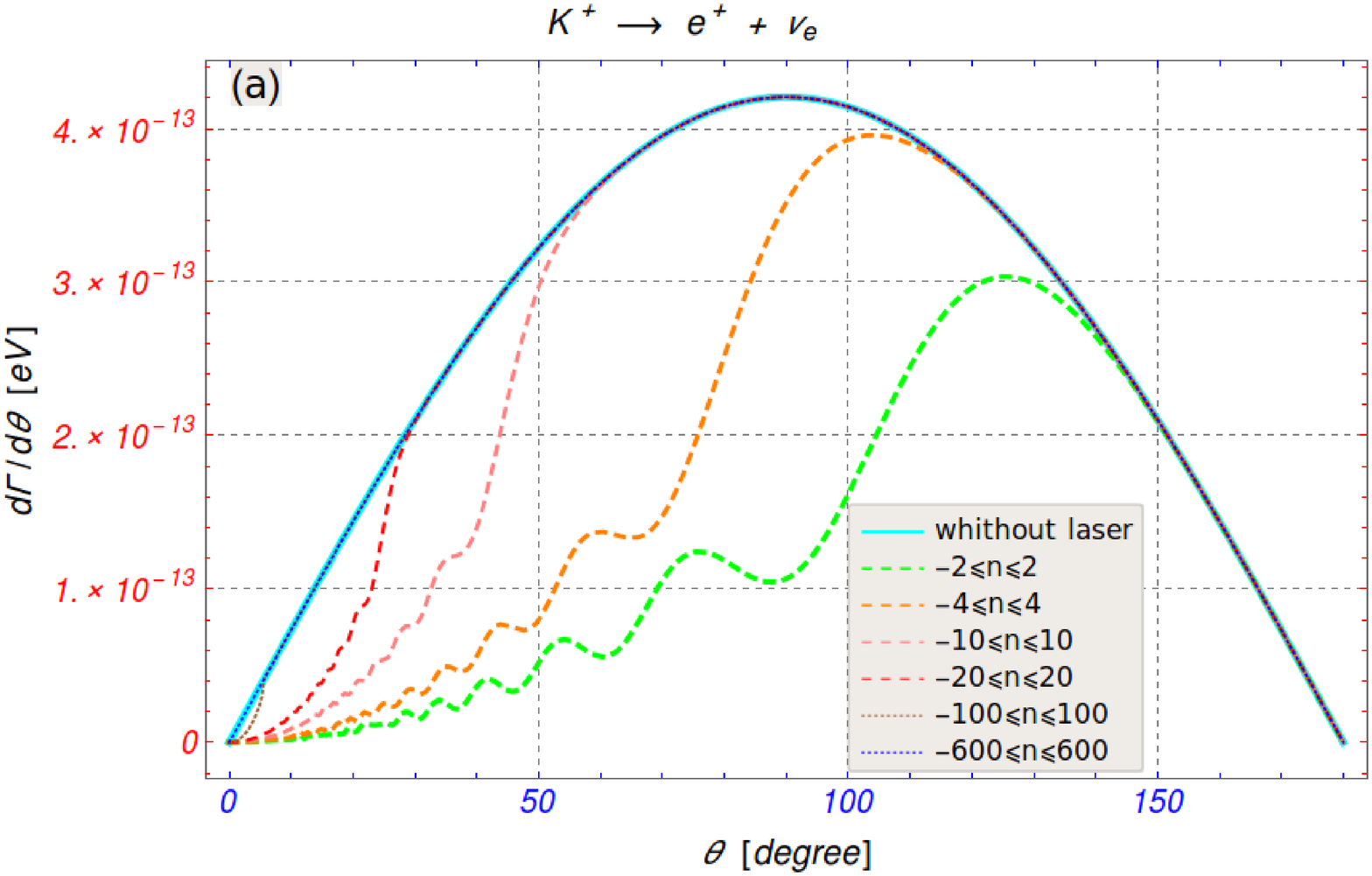}
  \end{minipage} %
  \hspace*{0.25cm}
  \begin{minipage}[t]{0.45\textwidth}
  \centering
    \includegraphics[width=\textwidth]{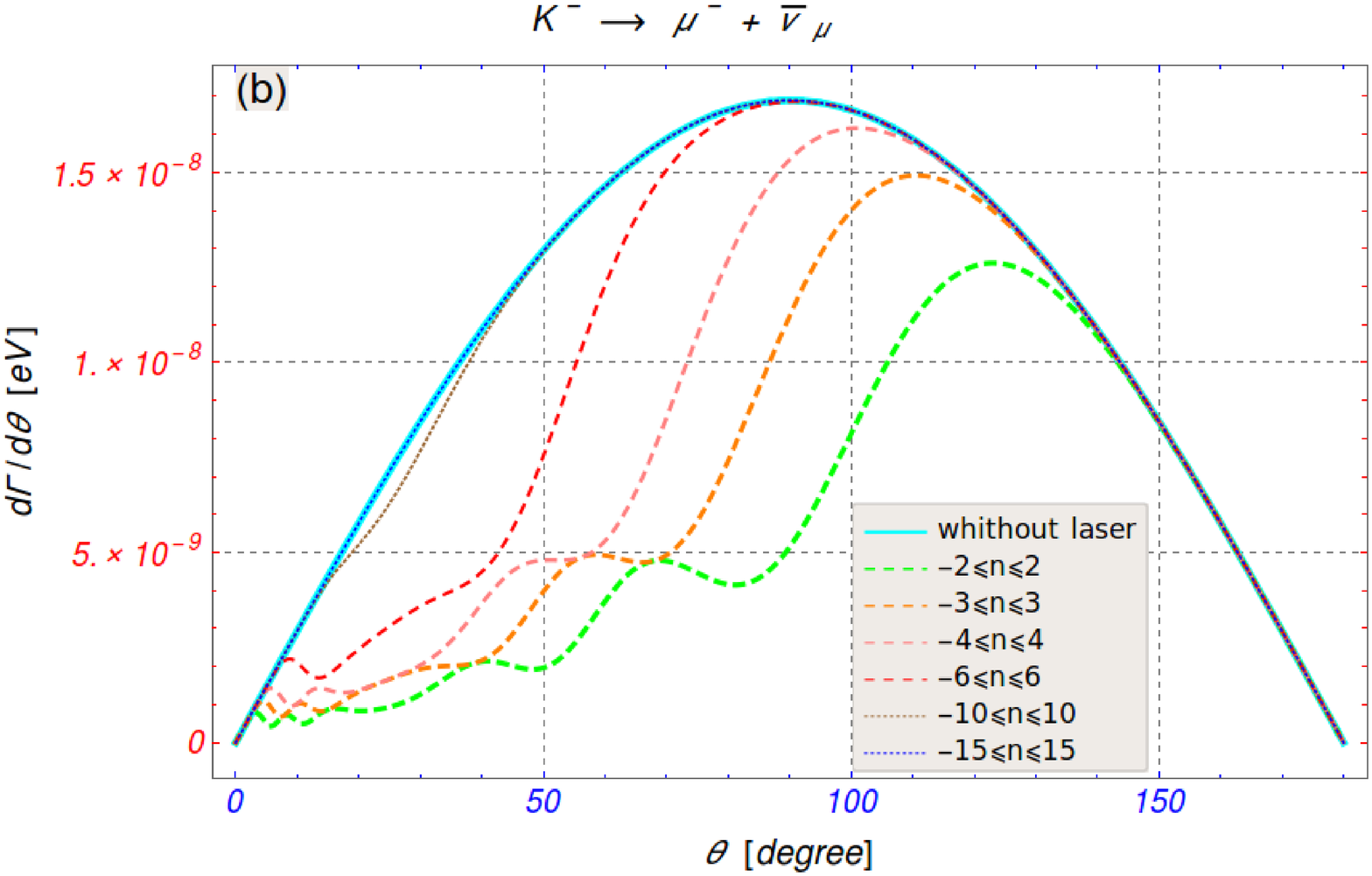}
  \end{minipage}
\caption{Dependence of the laser-assisted differential decay width of the charged kaon on the angle $\theta$ for different numbers of exchanged photons $n$. (a) represents the case of $K^{+}\rightarrow e^{+}+\nu_{e} $ decay. (b) represents the case of $K^{-}\rightarrow \mu^{-}+\bar{\nu}_{\mu}$ decay.}\label{Figure:2}
\end{figure}
Figure \ref{Figure:2} displays the effect of the number of exchanged photons $n$ on the differential decay width by varying the angle $ \theta $ and taking the laser parameters as $\hbar\omega=2\,eV$ and $\epsilon_{0}=10^{6}\,V/cm$. These two figures show that the circularly polarized laser field reduces the differential decay width by several orders of magnitudes. In addition, the differential decay width in the presence of the laser field converges to the laser-free differential decay width as much as we increase the number of exchanged photons $n$. Furthermore, when the number of exchanged photons reaches the cutoff which is the first value of $n$ for which \textbf{\textbf{d$\Gamma^{n}$/d$\theta$}} will be zero, the differential decay width of the charged kaon decay in the presence of the laser field will be equal to its corresponding laser-free differential decay width. This convegency is called the sum-rule shown by Bunkin and Fedorov, and also by Kroll and Watson\cite{Bunkin_and_Fedorov:1966}. It remains valid for any value of $n\geq $ cutoff. This result confirms that already obtained in figure \ref{Figure:1}(b). An other important point to mention here is that the order of magnitude of the exchanged photons number is different for the two studied processes. This is interpreted by the fact that the cutoff number increases as much as the mass of the produced charged lepton $l$ decreases. Let's move now to discuss the effect of the laser field on the two  leptonic decay channels of the antimatter $ K^{-} $ (\ref{partial_decay}) for different laser field strengths and frequencies.

\begin{table}[H]
\centering
\begin{tabular}{|p{1.5cm}|p{3cm}|p{3cm}|p{3cm}|p{3cm}|}
 \hline
     \multirow{2}{*}{} & \multicolumn{2}{c}{$\Gamma(K^{-}\rightarrow e^{-}+\bar{\nu}_{e} )\:\ [eV]$}\vline & \multicolumn{2}{c}{$\Gamma (K^{-}\rightarrow\mu^{-}+\bar{\nu}_{\mu})\:\ [eV]$} \vline \\
 \hline
  $\epsilon_{0}\:\ [\text{V/cm}]$ &~ {$\hbar\omega = 0.117\,eV$} &~ {$\hbar\omega =1.17\,eV$} &~ {$\hbar\omega = 0.117\,eV$} &~ {$\hbar\omega =1.17\,eV$} \\
 \hline
 ~~~~10 & ~~$8.41108 \times 10^{-13}$ &~~ $8.41108 \times 10^{-13}$ &~~ $3.37932 \times 10^{-8}$ &~~ $3.37932 \times 10^{-8}$ \\
~~~~$10^{2}$ &~~ $8.41108 \times 10^{-13}$ &~~ $8.41108 \times 10^{-13}$ &~~ $3.37932 \times 10^{-8}$ &~~ $3.37932 \times 10^{-8}$ \\
~~~~$10^{3}$ &~~ $8.38928 \times 10^{-13}$ &~~ $8.41108 \times 10^{-13}$ &~~ $3.37932 \times 10^{-8}$ &~~ $3.37932 \times 10^{-8}$ \\
~~~~$10^{4}$ &~~ $6.89024 \times 10^{-13}$ &~~ $8.41108 \times 10^{-13}$ &~~ $3.03290 \times 10^{-8}$ &~~ $3.37932 \times 10^{-8}$ \\
~~~~$10^{5}$ &~~ $1.18529 \times 10^{-13}$ &~~ $8.38928 \times 10^{-13}$ &~~ $5.24955 \times 10^{-9}$ &~~ $3.37932 \times 10^{-8}$ \\
~~~~$10^{6}$ &~~ $1.17859 \times 10^{-14}$ &~~ $6.89024 \times 10^{-14}$ &~~ $5.19151 \times 10^{-10}$ &~~ $3.03290 \times 10^{-8}$ \\
~~~~$10^{7}$ &~~ $1.17459 \times 10^{-15}$ &~~ $1.18529 \times 10^{-13}$ &~~ $5.18861 \times 10^{-11}$ &~~ $5.24955 \times 10^{-9}$ \\
~~~~$10^{8}$ &~~ $1.18045 \times 10^{-16}$ &~~ $1.17861 \times 10^{-14}$ &~~  $5.15933 \times 10^{-12}$ &~~ $5.19151 \times 10^{-10}$ \\
~~~~$10^{9}$ &~~ $1.33016 \times 10^{-17}$ &~~ $1.17586 \times 10^{-15}$ &~~ $5.27825 \times 10^{-13}$ &~~ $5.18861 \times 10^{-11}$ \\
~~~~$10^{10}$ &~~ $1.42073 \times 10^{-17}$ &~~ $1.30758 \times 10^{-16}$ &~~ $5.29958 \times 10^{-14}$ &~~ $5.15869 \times 10^{-12}$ \\
~~~~$10^{11}$ &~~ $1.30649 \times 10^{-16}$ &~~ $1.43220 \times 10^{-16}$ &~~ $5.38011 \times 10^{-15}$ &~~ $5.26224 \times 10^{-13}$ \\
~~~~$10^{12}$ &~~ $1.22483 \times 10^{-15}$ &~~ $1.30750 \times 10^{-15}$ &~~ $1.72866 \times 10^{-15}$ &~~ $5.39979 \times 10^{-14}$ \\
 \hline
\end{tabular}
  \caption{Laser-assisted partial decay width $ \Gamma $ of the two processes $K^{-}\rightarrow \mu^{-}+\bar{\nu}_{\mu}$  and $K^{-}\rightarrow e^{-}+\bar{\nu}_{e}$ for different values of $ \hbar\omega $ and $ \epsilon_{0} $ by taking  $n$ as ranging from $-20$ to $20$.}\label{Table:1}
\end{table}
Table \ref{Table:1} shows the effect of the circularly polarized laser field on the partial decay width of the two processes $K^{-}\rightarrow e^{-}+\bar{\nu}_{e}$ and $K^{-}\rightarrow \mu^{-}+\bar{\nu}_{\mu}$. We mention that, in the absence of an external electromagnetic field, the partial decay width of $ K^{-} $ in the electronic channel is $\Gamma^{-}=8.41108\times10^{-13}\,eV$ while $\Gamma^{-}=3.37932\times10^{-8}\,eV$ in the muonic channel \cite{PDG:2020}. This table indicates clearly that, for low intensities, the electromagnetic field doesn't affect the partial decay width regardless of the laser field frequency. For instance, for the process $K^{-}\rightarrow e^{-}+\bar{\nu}_{e}$, the effect of the laser field begins to appear at $ \epsilon_{0}\simeq 10^{3}\,V/cm$ for $\hbar\omega=0.117\,eV$. However, for $ \hbar\omega=1.17\,eV$, this effect appears at $\epsilon_{0}\simeq 10^{5}\,V/cm$. This result is explained by the fact that the number of exchanged photons $n=\pm 20 $ verifies the well known sum-rule for $ \epsilon_{0}\lesssim 10^{2}\,V/cm$ and $\hbar\omega=0.117\,eV$ as in the first column, or $ \epsilon_{0}\lesssim 10^{5}\,V/cm$  and $\hbar\omega=1.17\,eV$ as in the second column. To explain more clearly, we can deduce from figure \ref{Figure:2}(b) that, for $\hbar\omega=2\,eV$, $\epsilon_{0}=10^{6}\,V/cm$ and $n\geq 20$, the partial decay width in the muonic mode is equal to that obtained in the absence of the laser field. This is explained by the fact that there will be no photons to be exchanged for $n\geq 20$ (over limit of cutoff). Table \ref{Table:1} also shows that at high laser field strengths, the partial decay width of $K^{-}$ in the ($e^{-},\bar{\nu}_{e}$) channel approaches that in ($\mu^{-},\bar{\nu}_{\mu}$) channel. Therefore, to understand clearly this behavior, it is recommended to evaluate the effect of the circularly polarized laser field on the branching ratio. However, the fact that the charged kaon can decay via fifty modes, it is difficult to take into account all these modes to discuss the branching ratio and average lifetime. For this reason, we will discuss the laser-assisted leptonic branching ratio $ BR_{l} $ given by the equation (\ref{leptonic_BR}) and the charged kaon lifetime represented by the equation (\ref{lifetime}).

\begin{figure}[H]
\centering
  \begin{minipage}[t]{0.45\textwidth}
  \centering
    \includegraphics[width=\textwidth]{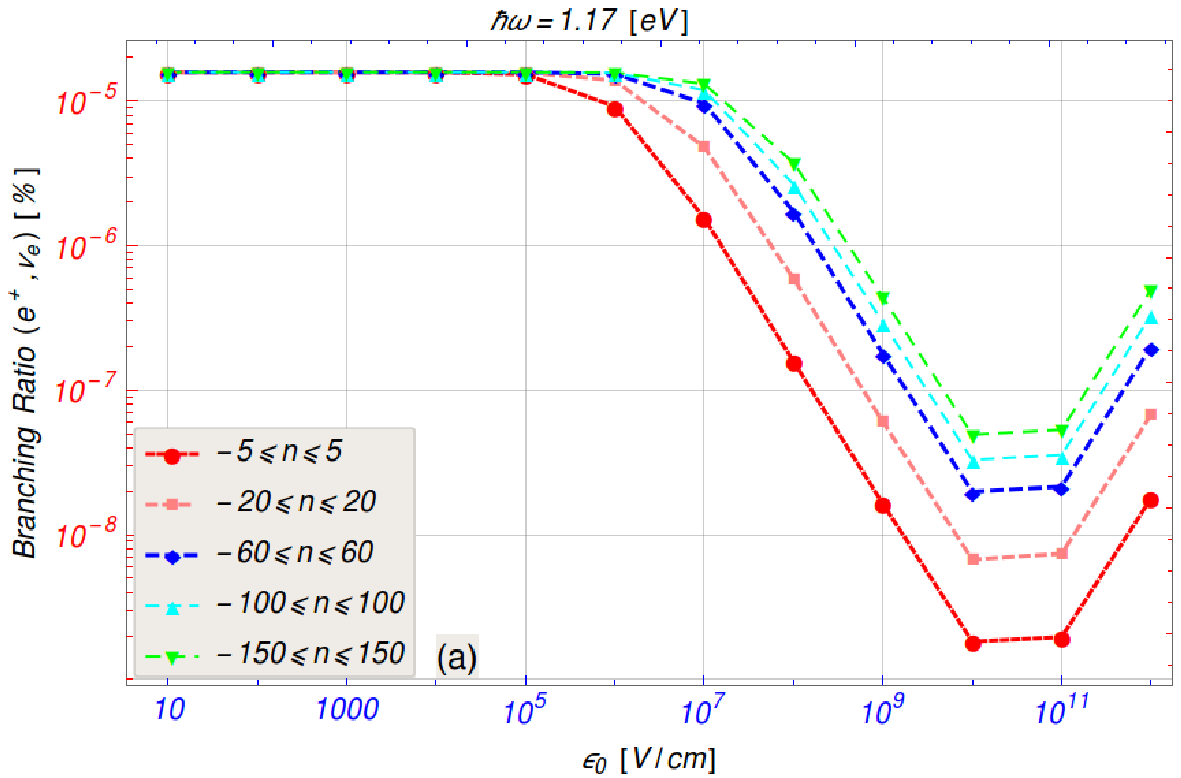}
  \end{minipage}
  \hspace*{0.25cm}
  \begin{minipage}[t]{0.45\textwidth}
  \centering
    \includegraphics[width=\textwidth]{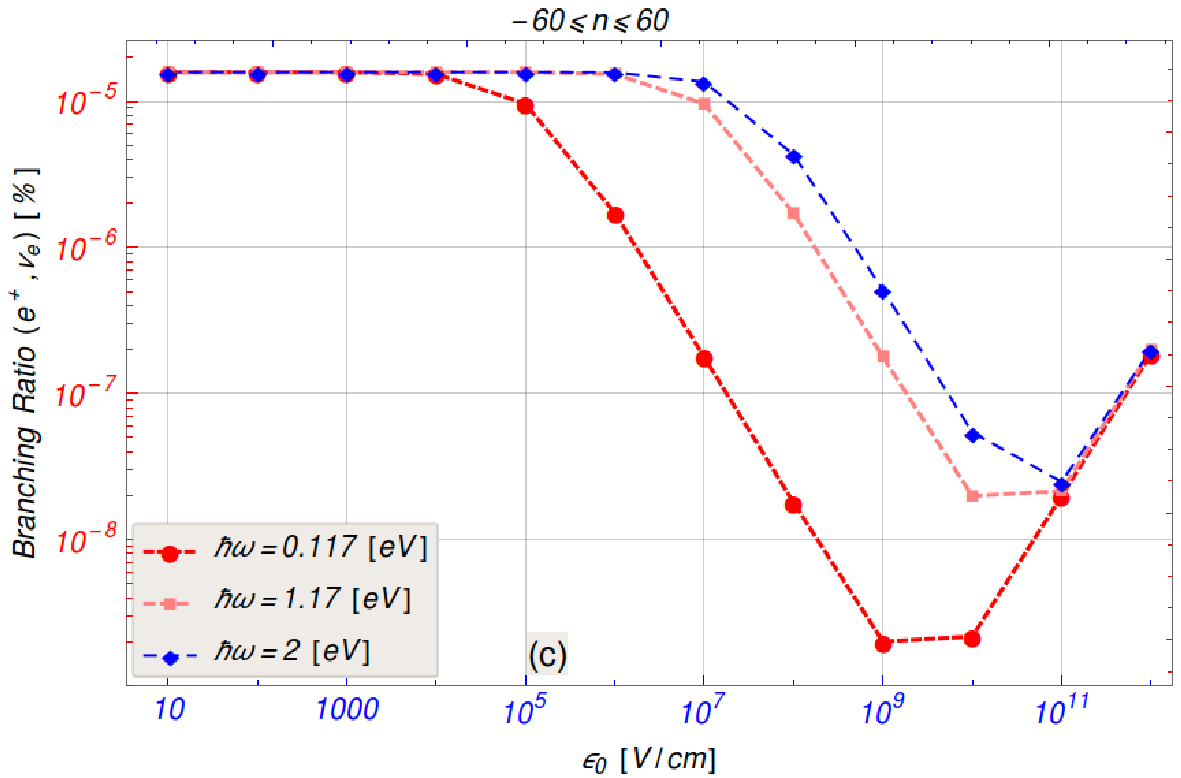}
  \end{minipage}
   \begin{minipage}[t]{0.45\textwidth}
  \centering
    \includegraphics[width=\textwidth]{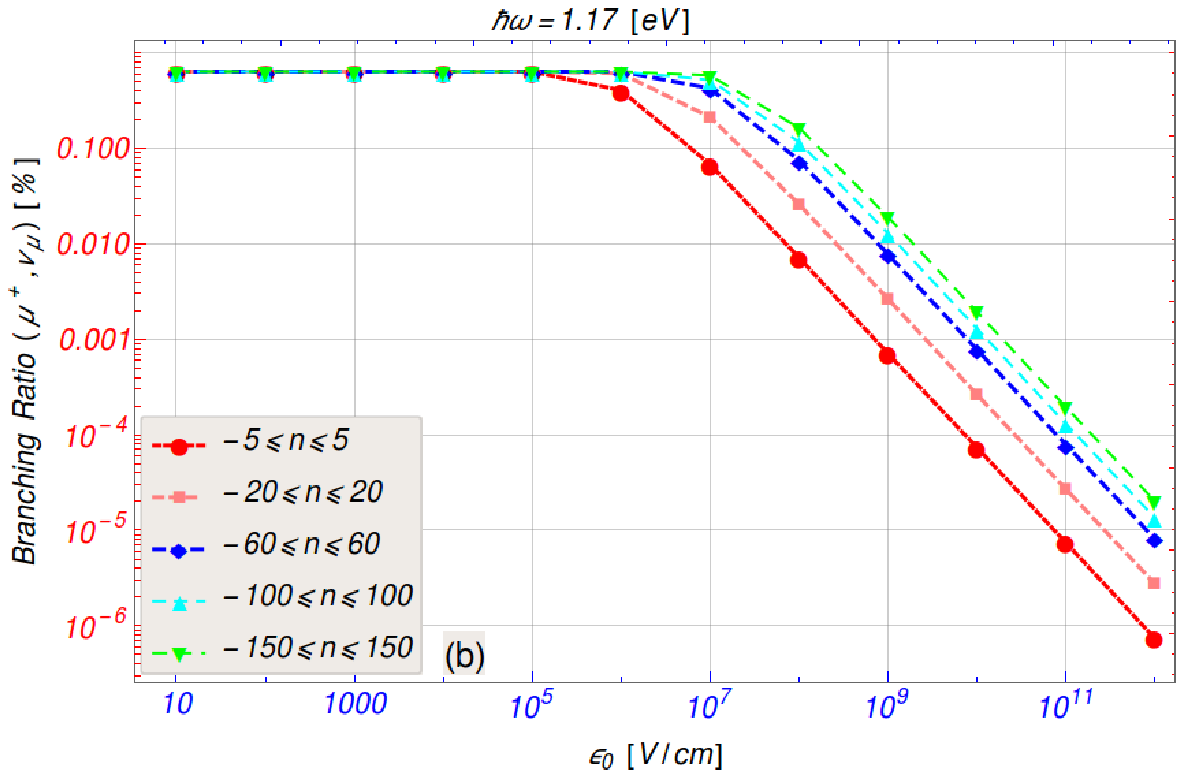}
  \end{minipage}
  \hspace*{0.25cm}
  \begin{minipage}[t]{0.45\textwidth}
  \centering
    \includegraphics[width=\textwidth]{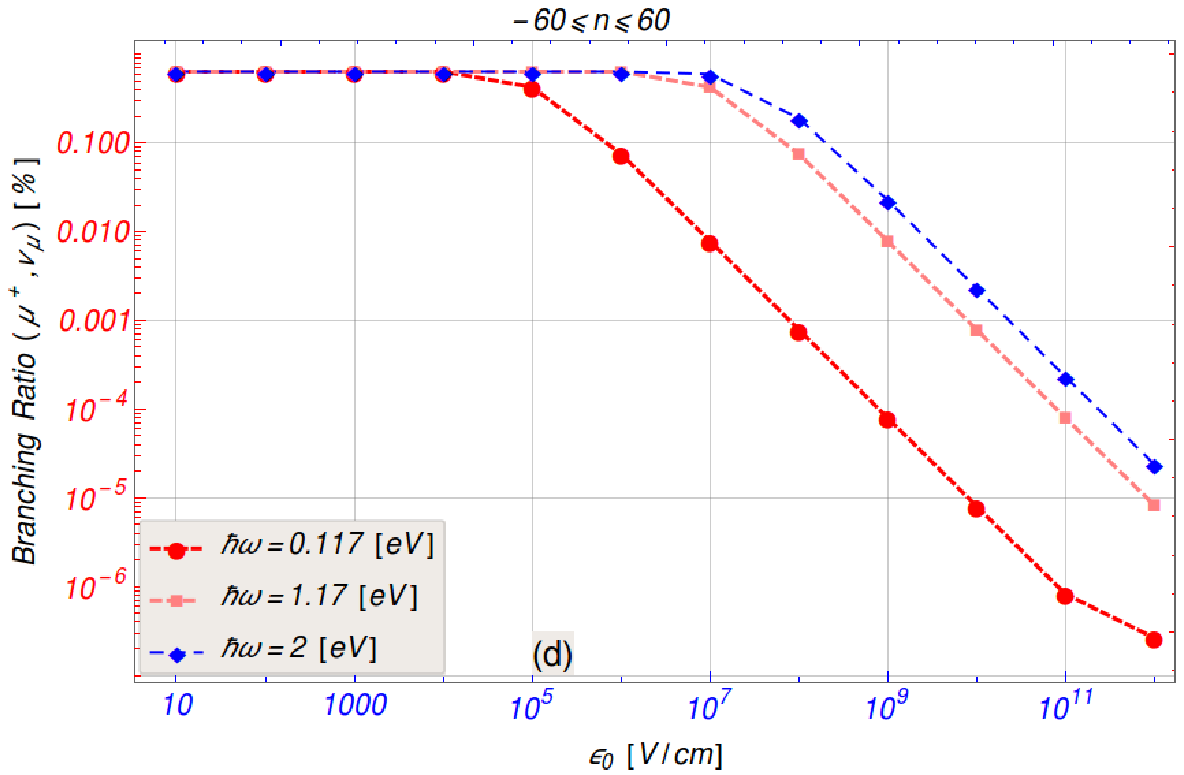}
  \end{minipage}
\caption{Laser-assisted leptonic branching ratio as a function of the laser field strength $\epsilon_{0}$. (a) and (b) are successively the leptonic branching ratios of $K^{+}\rightarrow e^{+}+\nu_{e}$ and $K^{+}\rightarrow \mu^{+}+\nu_{\mu}$ modes for different values of $n$ and for $ \hbar\omega=1.17\,eV$. (c) and (d) are the leptonic branching ratios of $K^{+}\rightarrow e^{+}+\nu_{e}$ and $K^{+}\rightarrow \mu^{+}+\nu_{\mu}$ modes, respectively, for different values of $ \hbar\omega $, and by summing over $n$ from $-60$ to $60$.}\label{Figure:3}
\end{figure}
Figure \ref{Figure:3} illustrates the dependence of the leptonic branching ratio $ BR_{l} $ of the positive kaon decay on the laser field strength for different number of exchanged photons and laser field frequencies. As we can see from figures \ref{Figure:3}(a) and \ref{Figure:3}(b), for $\hbar\omega=1.17 eV$, the variation of the $BR_{l}$ as a function of the laser field strength $\epsilon_{0}$ for different values of exchanged photons  $n$ ($ -j\leqslant n\leqslant j $, with $j=\{5,20,60,100,150\}$) indicates that, for low intensities, the laser field doesn't influence the leptonic branching ratio $BR_{l}$. Moreover, it starts to decrease in both channels when the laser field amplitude overcomes the threshold value $\epsilon_{0}=10^{5} V/cm$. However, as long as the laser field overcomes $\epsilon_{0}=10^{10}\,V/cm$, the branching ratio begins to increase in the electronic channel($e^{+},\nu_e$). It is well known that, in the absence of an external field, the experimental branching ratios of $ K^{+}\rightarrow \mu^{+}+\nu_{\mu}$ is $63.56(11)\%$ while the experimental branching ratio of $K^{+}\rightarrow e^{+}+\nu_{e}$ is equal to  $1.582(7)\times 10^{-5}$ \cite{PDG:2020}. Thus, we can deduce from Fig. \ref{Figure:3} that the circularly polarized laser field reduces the branching ratio of $K^{+}\rightarrow \mu^{+}+\nu_{\mu}$ and $K^{+}\rightarrow e^{+}+\nu_{e}$, however, the $BR(e^{+},\nu_{e})$ begins to increase at high laser field amplitudes. Figures \ref{Figure:3}(c) and \ref{Figure:3}(d) illustrate the dependence of the $BR_{l}$ as a function of the laser field strength by summing over the number of exchanged photons $n$ from -60 to 60 and for different known laser field frequencies which are $ \hbar\omega =0.117\,eV$, $ \hbar\omega =1.17\,eV$ and $ \hbar\omega =2\,eV$. It is obvious from these figures that, at high laser field strengths and regardless of the laser frequency, the leptonic branching ratio of the positive kaon decay always increases via the ($e^{+},\nu_e$) channel and keeps decreasing in the ($\mu^{+},\nu_{\mu}$) channel. In addition, when the laser field strength reaches a certain value, the two leptonic branching ratios become comparable. For instance, for $\hbar\omega=0.117\,eV$ and $\epsilon_{0}=10^{12}\,V/cm$, the leptonic branching ratios of the positive kaon decay modes are BR($e^{+},\nu_e)=1.86508\times10^{-7}$ and BR($\mu^{+},\nu_{\mu})=2.63299\times10^{-7}$. Furthermore, the variation of the leptonic branching ratio, regardless of the leptonic decay mode, is important as much as the laser frequency decreases. It is recognized that lifetime of an unstable particle is linked to its decay width, and since the leptonic decay width is affected by the laser field, it is worthy and important to study the effect of the external electromagnetic field on the lifetime $ \tau $ of the charged kaon as a function of the laser parameters.

\begin{figure}[H]
\centering
  \begin{minipage}[t]{0.45\textwidth}
  \centering
    \includegraphics[width=\textwidth]{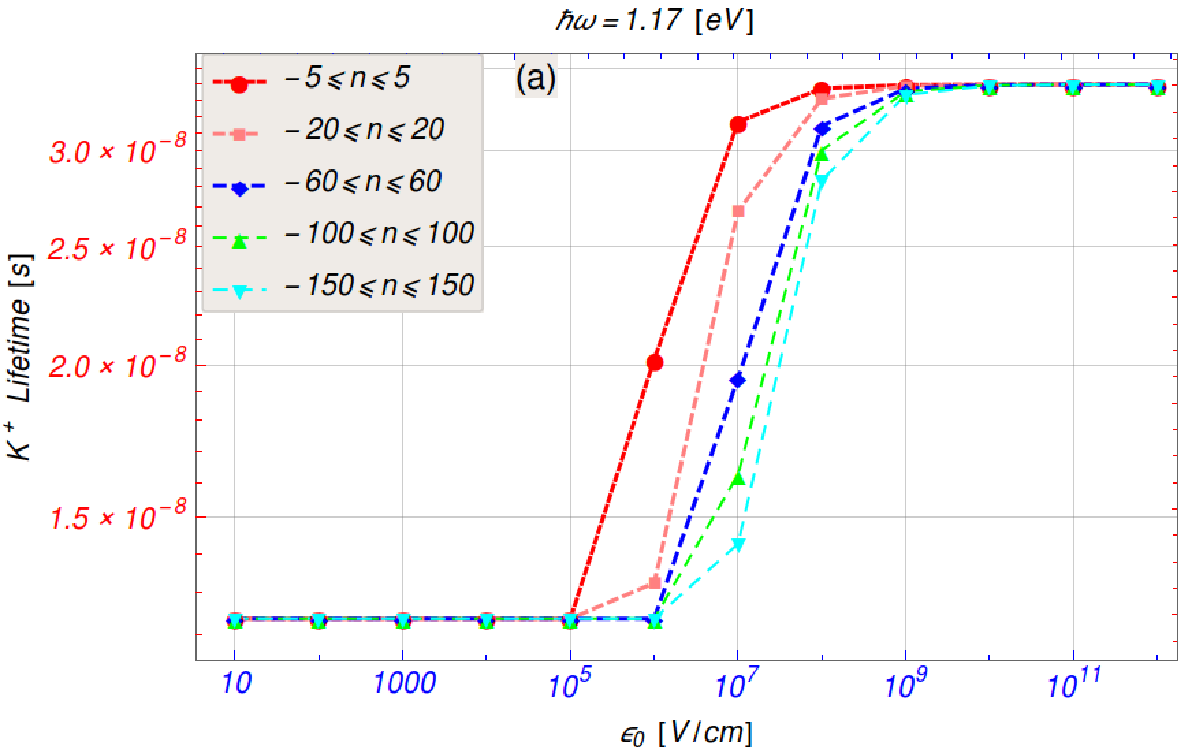}
  \end{minipage} %
  \hspace*{0.25cm}
  \begin{minipage}[t]{0.45\textwidth}
  \centering
    \includegraphics[width=\textwidth]{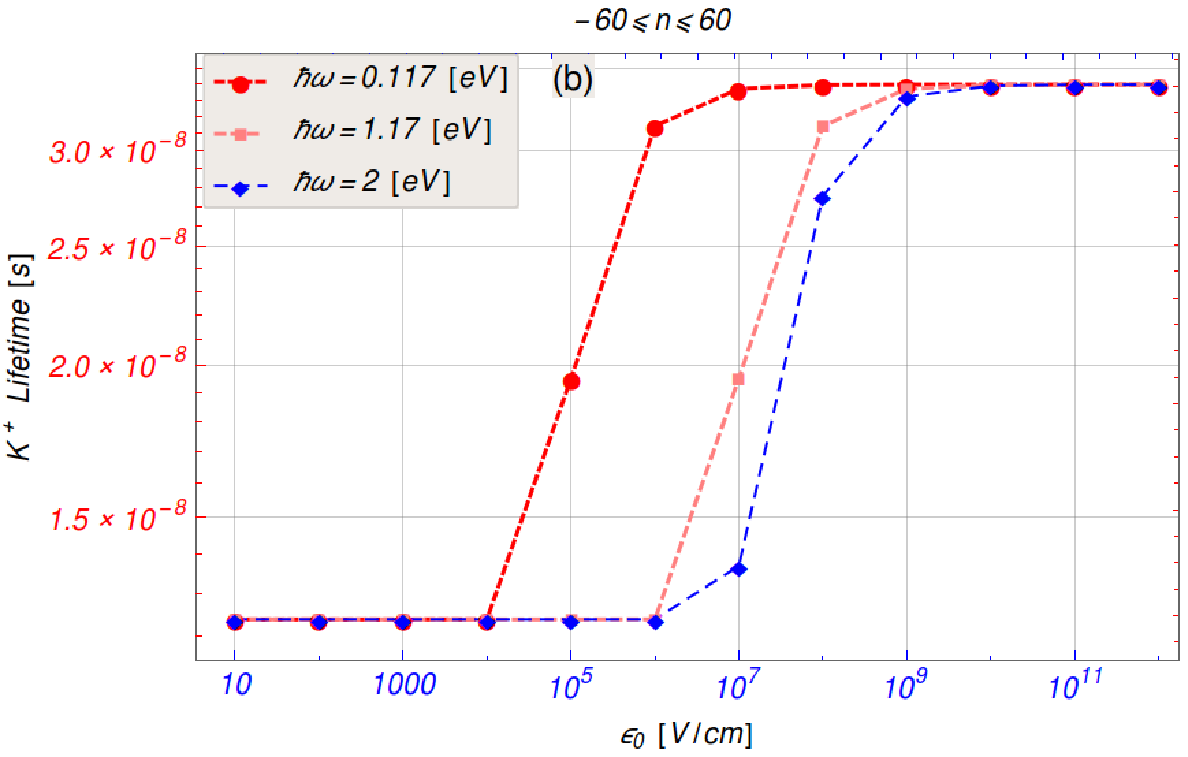}
  \end{minipage}
\caption{Lifetime of the positive kaon as a fucntion of the laser field strength. (a) the liftime of the positive kaon for different values of $n$ and for $ \hbar\omega=1.17\,eV$. (b) the liftime of the positive kaon for different known laser frequencies and by taking the number of exchanged photons as $-60\leq n\leq +60$.}\label{Figure:4}
\end{figure}
Figure \ref{Figure:4} displays the variation of the positive kaon lifetime $ \tau $ given by equation (\ref{lifetime}) on the laser field strength $\epsilon_{0}$ for different numbers of exchanged photons  $n$ and laser frequencies $\hbar\omega$. According to figure \ref{Figure:4}(a), for the laser frequency $\hbar\omega=1.17\,eV$ and regardless of the number of exchanged photons, the laser field does not affect the positive kaon lifetime unless its strength overcomes a threshold $\epsilon_{0}=10^5\,V/cm$. Whereas, for the number of exchanged photons $-60\leq n\leq +60$ as in figures \ref{Figure:4}(b), the laser field effect on the lifetime appears at different threshold values of the laser field strength which depends on the laser frequency. Furthermore, this threshold value decreases as much as the laser frequency decreases. Moreover, as $\epsilon_{0}$ overcomes this threshold value, the positive kaon lifetime increases as far as the laser field strength increases; it moves from the scale of $1.238 \times 10^{-8}\,s$ to $3.397 \times 10^{-8}\,s$. Therefore, the lifetime of the charged kaon is modified inside the electromagnetic field, despite the fact that we have only dressed its leptonic decay modes. To give a complete picture of the laser field effect on the leptonic decay of the charged kaon, we move to discuss the behavior of the leptonic ratio $R_{e/\mu}$ which is given by equation (\ref{leptonic_report}) inside the circularly polarized laser field.

\begin{figure}[H]
\centering
  \begin{minipage}[t]{0.45\textwidth}
  \centering
    \includegraphics[width=\textwidth]{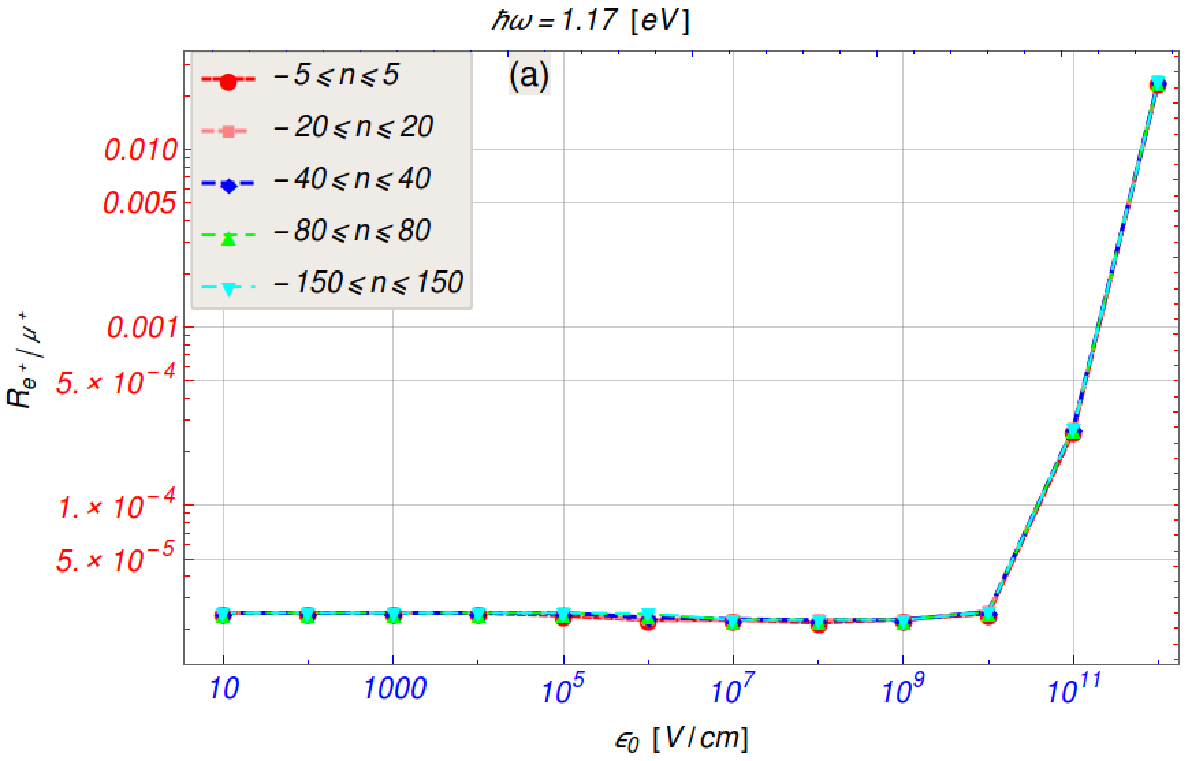}
  \end{minipage} %
  \hspace*{0.25cm}
  \begin{minipage}[t]{0.45\textwidth}
  \centering
    \includegraphics[width=\textwidth]{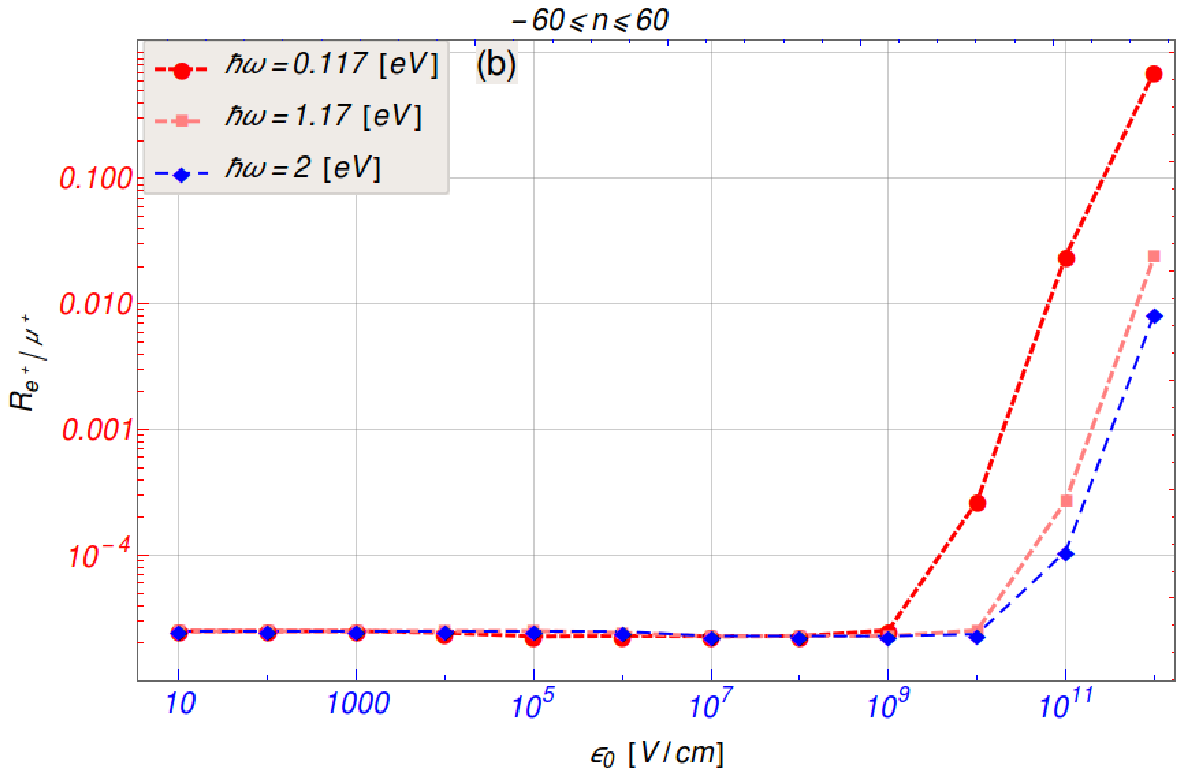}
  \end{minipage}
\caption{Leptonic ratio as a function of the laser field strength $\epsilon_{0}$. (a): $R_{e/\mu}$ for different values of $n$ and $\hbar\omega=1.17\,eV$. (b): $R_{e/\mu}$ for different values of $ \hbar\omega$ and by summing over $n$ from $-60$ to $60$.}\label{Figure:5}
\end{figure}
Figure \ref{Figure:5} displays the variation of the leptonic ratio $R_{e/\mu}$ as a function of the laser field strength for different number of exchanged photons by summing over $n$ from $-i$ to $i$ such that $i=\{5,20,40,80,150\}$ as in figure \ref{Figure:5}(a), and for different laser frequencies $\hbar\omega$ as in figure \ref{Figure:5}(b). We mention that, in the absence of an external field, the quantity $R_{e/\mu}$ is experimentally equal to $2.488(9)\times10^{-5}$. Thus, we notice that the effect of the laser field on the experimental factor $R_{e/\mu}$ becomes important when the intensity $\epsilon_{0}$ is large, and that confirms the significant decrease of the charged kaon decay width in the muonic channel in comparison with that found in the electronic mode. For example, for $ \hbar\omega=0.117\,eV$ and $\epsilon_{0}=10^{12}\,V/cm$, we can deduce from figure \ref{Figure:5}(a) that $\Gamma (K^{+}\rightarrow e^{+}+\nu_{e})= 0.708347\times\Gamma (K^{+}\rightarrow \mu^{+}+\nu_{\mu})$. From this result, we can confirm that, in the presence of the laser field and by dressing all the decay modes of the charged kaon, the branching ratio in the electronic channel will be of the same importance as the branching ratio in the muonic channel. The last important point to be discussed in this research paper is the variation of the quantity associated to CPT symmetry (\ref{CPT_parameter}). It is an experimental quantity which represents one of the most challenging issues that are not explained by the standard model.

\begin{table}[H]
 \centering
\begin{tabular}{|p{1.5cm}|p{3cm}|p{3cm}|p{3cm}|p{3cm}|}
 \hline
     \multirow{2}{*}{} & \multicolumn{2}{c}{$\Delta_{CPT} (e,\nu_{e})$} \vline & \multicolumn{2}{c}{$\Delta_{CPT} (\mu,\nu_{\mu})$} \vline\\

 \hline
  $\epsilon_{0}\:\ [\text{V/cm}]$ &~ {$\hbar\omega = 0.117\,eV$} &~ {$\hbar\omega =1.17\,eV$} &~ {$\hbar\omega = 0.117\,eV$} &~ {$\hbar\omega =1.17\,eV$} \\
 \hline
 ~~~~  10 & $-1.24871\times10^{-11}$ &~ $5.53546\times10^{-12}$ &~ $9.79109\times10^{-17}$ &~ $9.79109\times10^{-17}$ \\
~~~~$10^{2}$ & $-4.72363\times10^{-12}$ & $-2.51076\times10^{-12}$ & $-3.91644\times10^{-16}$ & $-2.93733\times10^{-16}$ \\
~~~~$10^{3}$ & $-1.08599\times10^{-11}$ &~ $2.42769\times10^{-11}$ & $-4.89555\times10^{-16}$ &~ $1.95822\times10^{-16}$ \\
~~~~$10^{4}$ &~ $5.80142\times10^{-11}$ & $-3.06077\times10^{-12}$ &~ $8.32847\times10^{-11}$ &~~~~~~~~~ $0$ \\
~~~~$10^{5}$ &~ $1.59613\times10^{-10}$ &~ $1.27065\times10^{-11}$ &~ $2.56883\times10^{-10}$ & $-1.95822\times10^{-16}$ \\
~~~~$10^{6}$ &~ $9.85136\times10^{-10}$ & $-5.24162\times10^{-11}$ &~ $1.11873\times10^{-9}$ &~ $8.32833\times10^{-10}$ \\
~~~~$10^{7}$ &~ $8.25115\times10^{-9}$ & $-3.49834\times10^{-10}$ &~ $1.57303\times10^{-9}$ &~ $2.56878\times10^{-9}$ \\
~~~~$10^{8}$ &~ $1.17815\times10^{-7}$ &~ $7.86274\times10^{-9}$ &~ $3.28896\times10^{-8}$ &~ $1.11873\times10^{-8}$ \\
~~~~$10^{9}$ & $-8.78207\times10^{-7}$ &~ $8.16060\times10^{-8}$ &~ $1.09299\times10^{-6}$ &~ $1.57062\times10^{-8}$ \\
~~~~$10^{10}$ & $-1.36747\times10^{-8}$ &~ $5.59991\times10^{-7}$ &~ $2.25762\times10^{-6}$ &~ $3.35539\times10^{-7}$ \\
~~~~$10^{11}$ & $-6.10246\times10^{-5}$ &~ $7.12923\times10^{-7}$ &~ $1.82039\times10^{-4}$ & $-4.19000\times10^{-6}$ \\
~~~~$10^{12}$ &~ $7.56812\times10^{-4}$ & $-8.71563\times10^{-5}$ &~ $5.54973\times10^{-4}$ & $-2.13747\times10^{-4}$ \\
\hline
  \end{tabular}
 \caption{$\Delta_{CPT}$ as a function of the laser field strength $ \epsilon_{0} $ for different laser field frequencies $\hbar\omega $.}\label{Table:2}
\end{table}
Table \ref{Table:2} represents the variation of the parameter associated to CPT symmetry, which is given by equation (\ref{CPT_parameter}), as a function of the laser field strength $\epsilon_{0}$ for different values of $ \hbar\omega $ and by summing over the exchanged photons number $n$ form $-20$ to $+20$. As we can see from the table \ref{Table:2}, it is clear that the circularly polarized laser field affects both matter and antimatter decay of the charged kaon. Additionally, it is recognized that CPT symmetry is preserved ($\vartriangle_{CPT}=0\%$) or broken ($\vartriangle_{CPT}\neq 0\%$) in both cases where the disintegrated particles $K^{\pm}$ have the same or different masses. According to PDG \cite{PDG:2020}, the experimental value of the CPT associated parameter in the muonic mode is of the order of $(-0.27 \pm 0.21)\%$. However, in the absence of the laser field, its theoretical value is equal to zero (equation (\ref{decay_without_laser})) while it goes from an order of $10^{-17}$ to $10^{-4}$ when we increase the electric field strength form $\epsilon_{0}=10\,V/cm$ to $\epsilon_{0}=10^{12}\,V/cm$. Moreover, for the positive values of $\vartriangle_{CPT}$, we see that the matter $K^{+}$ becomes dominant at high precision width value $\Gamma$ in comparison to the antimatter $K^{-}$ while its negative values indiquate the dominance of the antimatter over the matter. CPT symmetry is also affected by the external field in the electronic channel, and the dominance of matter over antimatter or vice-versa depends of the laser field parameters. The results obtained about CPT conservation and CPT symmetry breaking are due to the fact that phases associated to the Bessel functions which are expressed as a function of $\phi_{0}$ in equation (\ref{Bessel_transformations}) depends on the type of disintegration (matter or antimatter). Furthermore, the elements associated to the lepton mass $m_{l}$ which is obtained from the Dirac bispinors part are not equal to zero as in the case of the laser-free disintegration. All these results of CPT breaking symmetry can represent, in cosmology, a probable reason for the asymmetry observed in the universe between matter and antimatter \cite{Sakharov:1967}.
\section{Conclusion}\label{Sec3}
In this study, we have investigated the laser-assisted leptonic decay of the charged kaon at the lowest order. We have found that the circularly polarized laser field reduces the differential decay width by several orders of magnitudes. In addition, at high laser field intensities, the partial decay width of the negative kaon ($k^{-}$) in the ($e^{-}$,$\nu_{e}$) channel approaches that in $(\mu^{-},\nu_{\mu})$. Moreover, the branching ratio of the two processes $ K^{+} \rightarrow e^{+}+\nu_{e}$ and $ K^{+} \rightarrow\mu^{+}+\nu_{\mu}$ decreases in the presence of an external field. However, the branching ratio of $ K^{+} \rightarrow e^{+}+\nu_{e}$ begins to increases when the laser field strength reaches $\epsilon_{0}=10^{10}\,V/cm$. Since lifetime is inversely proportional to the total decay width, which is affected by the laser field despite the fact that we have only dressed the leptonic modes, we assume that lifetime of the charged kaon extends. Furthermore, the variation of the leptonic ratio $R_{e/\mu}$ as a function of the laser field strength indicates that the two leptonic decay widths of the the charged kaon will be of the same importance at high laser field intensities. By theoretically measuring the  parameter associated to CPT symmetry inside the electromagnetic field, we have come out with a surprising result which indicates that we can control the charged kaon decay by applying an external field to conserve or violate the CPT symmetry. Consequently, by violating this symmetry, we have the possibility to make the antimatter dominant over the matter or vice versa. Additionally, the laser field also enables us to conserve this symmetry in order to produce the same amount of matter and antimatter.
\appendix
\section*{Appendix}
The integral over $ \vert\vec{q}_{2}\vert $ in the equation (\ref{partial_decay}) can be evaluated by using the following mathematical formula:
\begin{eqnarray}
\int F(x)\delta(G(x))dx=\dfrac{F(x_{0})}{|G'(x_{0})|} \qquad \text{where} \quad G(x_{0})=0,
\end{eqnarray}
and so, our next integral becomes:
\begin{eqnarray}
\int_{0}^{+\infty} \dfrac{\vec{q}_{2}^{2}d\vert\vec{q}_{2}\vert}{Q_{2}E_{3}}\delta(Q_{1}-Q_{2}-E_{3}+nw)\displaystyle{\sum_{s_{2},s_{3}}}\vert \mathcal{M}_{fi}^{n\pm}\vert^{2}=\dfrac{\displaystyle{\vec{\textbf{q}}_{2}^{2}\sum_{s_{2},s_{3}}}\vert \mathcal{M}_{fi}^{n\pm}\vert^{2}}{Q_{2}E_{3}|G'(\vert\vec{\textbf{q}}_{2}\vert)|}
\end{eqnarray}
where $ Q_{2}=\sqrt{\vec{\textbf{q}}_{2}^{2}+m_{l}^{*2}} $ and $E_{3}=\sqrt{\vec{P}_{3}^{2}}=|\vec{q}_{1}-\vec{q}_{2}+n\vec{k}|$. The absolute value of $ G'(\vert\vec{\textbf{q}}_{2}\vert) $ is deduced from the first derivation with respect to $ \vert\vec{\textbf{q}}_{2}\vert $. Thus ,it becomes as follows:
\begin{eqnarray}
G'(|\vec{\textbf{q}}_{2}|)&=&-\dfrac{|\vec{\textbf{q}}_{2}|}{\sqrt{|\vec{\textbf{q}}_{2}|^{2}+\omega^{2}-e^{2}a^{2}}}-\left[ |\vec{\textbf{q}}_{2}|-\omega\cos\theta(n-\dfrac{e^{2}a^{2}}{2k.P_{1}})\right] \nonumber \\
&\times&\dfrac{1}{\sqrt{|\vec{\textbf{q}}_{2}|^{2}-2|\vec{\textbf{q}}_{2}|\omega\cos\theta(n-\dfrac{e^{2}a^{2}}{2k.P_{1}})+\omega^{2}(n-\dfrac{e^{2}a^{2}}{2k.P_{1}})^{2}}},
\end{eqnarray}
where the four-vector $\vec{\textbf{q}}_{2}$ has the following components $ \vec{\textbf{q}}_{2}\equiv |\vec{\textbf{q}}_{2}|(\sin\theta\cos\varphi,\sin\sin,\cos\theta)$ is in spherical coordinates. Its norm $|\vec{\textbf{q}}_{2}|$ is given by:
\begin{eqnarray}
|\vec{\textbf{q}}_{2}|&=&(\omega(e^{2}a^{2}-2nk.P_{1})\times(e^{4}a^{4}\omega^{2}-4e^{2}a^{2}n\omega^{2}k.P_{1}-4k.P_{1} \nonumber \\
&\times &(m^{*2}_{l}+(E_{1}-\dfrac{e^{2}a^{2}}{2k.P_{1}}w)^{2}+2(E_{1}-\dfrac{e^{2}a^{2}}{2k.P_{1}}w)\omega))\cos\theta\nonumber\\
&+&2k.P_{1}\sqrt{((E_{1}-\dfrac{e^{2}a^{2}}{2k.P_{1}}w)+n\omega)^{2}}(a^{8}e^{8}\omega^{4}-8a^{6}e^{6}k.P_{1}n\omega^{4}\nonumber\\
&+&32a^{2}e^{2}(k.P_{1})^{3}(E_{1}-\dfrac{e^{2}a^{2}}{2k.P_{1}}w)n\omega^{2}(E_{1}-\dfrac{e^{2}a^{2}}{2k.P_{1}}w+2n\omega)\nonumber\\
&-&8a^{4}e^{4}(k.P_{1})^{2}\omega^{2}((E_{1}-\dfrac{e^{2}a^{2}}{2k.P_{1}}w)^{2}+2(E_{1}-\dfrac{e^{2}a^{2}}{2k.P_{1}}w)n\omega \nonumber\\
&-&2s^{2}\omega^{2})+16(k.P_{1})^{4}(m^{*4}_{l}-2m^{*2}_{l}(E_{1}-\dfrac{e^{2}a^{2}}{2k.P_{1}}w+n\omega)^{2}\nonumber\\
&+&(E_{1}-\dfrac{e^{2}a^{2}}{2k.P_{1}}w)^{2}(E_{1}-\dfrac{e^{2}a^{2}}{2k.P_{1}}w+2n\omega)^{2}\nonumber)\\
&+&8(k.P_{1})^{2}m^{*2}_{l}(a^{2}e^{2}-2nk.P_{1})^{2}\omega^{2}\cos2\theta)^{1/2})\nonumber\\
&\times &\dfrac{1}{(k.P_{1})^{3}(E_{1}-\dfrac{e^{2}a^{2}}{2k.P_{1}}w+n\omega)^{2}-4k.P_{1}(a^{2}e^{2}-2nk.P_{1})^{2}\omega^{2}\cos^{2}\theta}.
\end{eqnarray}
To give the expression of the quantity $\sum_{s_{2},s_{3}}\vert {{\mathcal{M}}_{fi}}^{n\pm} \vert^{2}$ in equation (\ref{partial_decay}), we have used the FeynCalc-9.3.0 package. Thus, we have:
\begin{eqnarray}
&\sum_{s_{2},s_{3}}&|{{\mathcal{M}}_{fi}}^{n\pm}|^{2}=\mp 4em_{K}^{2}(a_{1}.P_{3})\boldsymbol{B}1n^{\pm}(z)\boldsymbol{B}n^{\pm}(z)^{*}\mp 4em_{K}^{2}(a_{2}.P_{3})\boldsymbol{B}2n^{\pm}(z)\boldsymbol{B}n^{\pm}(z)^{*} \nonumber \\
&\mp &4em_{K}^{2}(a_{1}.P_{3})\boldsymbol{B}1n^{\pm}(z)^{*}\boldsymbol{B}n^{\pm}(z)\mp 4em_{K}^{2}(a_{2}.P_{3})\boldsymbol{B}2n^{\pm}(z)^{*}\boldsymbol{B}n^{\pm}(z) \nonumber \\
&+&\varepsilon[a_{1},a_{2},k,P_{3}]\left[ \dfrac{4ie^{2}m_{K}^{2}\boldsymbol{B}2n^{\pm}(z)\boldsymbol{B}1n^{\pm}(z)^{*}-4ie^{2}m_{K}^{2}\boldsymbol{B}1n^{\pm}(z)\boldsymbol{B}2n^{\pm}(z)^{*}}{k.P_{2}} \right] \nonumber \\
&\pm &\varepsilon[a_{1},k,P_{2},P_{3}]\left[ \dfrac{4iem_{K}^{2}\boldsymbol{B}1n^{\pm}(z)\boldsymbol{B}n^{\pm}(z)^{*}-4iem_{K}^{2}\boldsymbol{B}1n^{\pm}(z)^{*}\boldsymbol{B}n^{\pm}(z)}{k.P_{2}} \right] \nonumber \\
&\pm &\varepsilon[a_{2},k,P_{2},P_{3}]\left[ \dfrac{4iem_{K}^{2}\boldsymbol{B}2n^{\pm}(z)\boldsymbol{B}n^{\pm}(z)^{*}-4iem_{K}^{2}\boldsymbol{B}2n^{\pm}(z)^{*}\boldsymbol{B}n^{\pm}(z)}{k.P_{2}} \right] \nonumber \\
&+&\varepsilon[a_{1},a_{2},k,P_{1}]\left[ \dfrac{8ie^{2}(P_{1}.P_{3})\boldsymbol{B}1n^{\pm}(z)\boldsymbol{B}2n^{\pm}(z)^{*}-8ie^{2}(P_{1}.P_{3})\boldsymbol{B}2n^{\pm}(z)\boldsymbol{B}1n^{\pm}(z)^{*}}{k.P_{2}} \right] \nonumber \\
&\pm &\varepsilon[a_{1},k,P_{1},P_{2}]\left[ \dfrac{8ie(P_{1}.P_{3})\boldsymbol{B}1n^{\pm}(z)\boldsymbol{B}n^{\pm}(z)^{*}-8ie(P_{1}.P_{3})\boldsymbol{B}n^{\pm}(z)\boldsymbol{B}1n^{\pm}(z)^{*}}{k.P_{2}} \right] \nonumber \\
&\pm &\varepsilon[a_{2},k,P_{1},P_{2}]\left[ \dfrac{8ie^{2}(P_{1}.P_{3})\boldsymbol{B}2n^{\pm}(z)\boldsymbol{B}n^{\pm}(z)^{*}-8ie(P_{1}.P_{3})\boldsymbol{B}n^{\pm}(z)\boldsymbol{B}2n^{\pm}(z)^{*}}{k.P_{2}} \right] \nonumber \\
&+&\left[ \dfrac{4e^{2}m_{K}^{2}a^{2}(k.P_{3})\boldsymbol{B}1n^{\pm}(z)\boldsymbol{B}1n^{\pm}(z)^{*}+4e^{2}m_{K}^{2}a^{2}(k.P_{3})\boldsymbol{B}2n^{\pm}(z)\boldsymbol{B}2n^{\pm}(z)^{*}}{k.P_{2}} \right] \nonumber \\
&\pm &\left[ \dfrac{4em_{K}^{2}(a_{1}.P_{2})(k.P_{3})\boldsymbol{B}1n^{\pm}(z)\boldsymbol{B}n^{\pm}(z)^{*}+4em_{K}^{2}(a_{2}.P_{2})(k.P_{3})\boldsymbol{B}2n^{\pm}(z)\boldsymbol{B}n^{\pm}(z)^{*}}{k.P_{2}} \right] \nonumber \\
&\pm &\left[ \dfrac{4em_{K}^{2}(a_{1}.P_{2})(k.P_{3})\boldsymbol{B}1n^{\pm}(z)^{*}\boldsymbol{B}n^{\pm}(z)+4em_{K}^{2}(a_{2}.P_{2})(k.P_{3})\boldsymbol{B}2n^{\pm}(z)^{*}\boldsymbol{B}n^{\pm}(z)}{k.P_{2}} \right] \nonumber \\
&-&\left[ \dfrac{8e^{2}a^{2}(P_{1}.P_{3})(k.P_{1})\boldsymbol{B}1n^{\pm}(z)\boldsymbol{B}1n^{\pm}(z)^{*}+8e^{2}a^{2}(P_{1}.P_{3})(k.P_{1})\boldsymbol{B}2n^{\pm}(z)\boldsymbol{B}2n^{\pm}(z)^{*}}{k.P_{2}} \right] \nonumber \\
&\mp &\left[ \dfrac{8e(a_{1}.P_{2})(P_{1}.P_{3})(k.P_{1})\boldsymbol{B}1n^{\pm}(z)\boldsymbol{B}n^{\pm}(z)^{*}+8e(a_{2}.P_{2})(P_{1}.P_{3})(k.P_{1})\boldsymbol{B}2n^{\pm}(z)\boldsymbol{B}n^{\pm}(z)^{*}}{k.P_{2}} \right] \nonumber \\
&\mp &\left[ \dfrac{8e(a_{1}.P_{2})(P_{1}.P_{3})(k.P_{1})\boldsymbol{B}1n^{\pm}(z)^{*}\boldsymbol{B}n^{\pm}(z)+8e(a_{2}.P_{2})(P_{1}.P_{3})(k.P_{1})\boldsymbol{B}2n^{\pm}(z)^{*}\boldsymbol{B}n^{\pm}(z)}{k.P_{2}} \right] \nonumber \\
&+&\left[ 16(P_{1}.P_{2})(P_{1}.P_{3})\boldsymbol{B}n^{\pm}(z)\boldsymbol{B}n^{\pm}(z)^{*} \right]-\left[ 8m_{K}^{2}(P_{2}.P_{3})\boldsymbol{B}n^{\pm}(z)\boldsymbol{B}n^{\pm}(z)^{*} \right],
\end{eqnarray}
We have used Grozin's convention, $ \varepsilon_{0123}=1 $, to calculate the terms $ \varepsilon[A,B,C,D] $ that appear during the evaluation of the quantity $ \sum_{s_{2},s_{3}}|{{\mathcal{M}}_{fi}}^{n\pm}|^{2}$. These terms are calculated by using the following calculation techniques:
\begin{eqnarray}
\varepsilon[a_{1},a_{2},k,P_{3}]&=&\varepsilon_{1203}a_{1}^{1}a_{2}^{2}k^{0}P_{3}^{3}+\varepsilon_{1230}a_{1}^{1}a_{2}^{2}k^{3}P_{3}^{0},\\
&=&\dfrac{\epsilon_{0}^{2}}{\omega}\lbrace\left[\omega(n-\dfrac{e^{2}a^{2}}{2k.P_{1}})-|\vec{\textbf{q}}_{2}|\cos\theta \right] \nonumber\\
&-&\sqrt{|\vec{\textbf{q}}_{2}|^{2}-2|\vec{\textbf{q}}_{2}|\omega\cos\theta(n-\dfrac{e^{2}a^{2}}{2k.P_{1}})+\omega^{2}(n-\dfrac{e^{2}a^{2}}{2k.P_{1}})^{2}}\;\ \rbrace,\\
\varepsilon[a_{1},k,P_{2},P{3}]&=&\varepsilon_{1023}a_{1}^{1}k^{0}P_{2}^{2}P_{3}^{3}+\varepsilon_{1032}a_{1}^{1}k^{0}P_{2}^{3}P_{3}^{2}+\varepsilon_{1302}a_{1}^{1}k^{3}P_{2}^{0}P_{3}^{2}+\varepsilon_{1320}a_{1}^{1}k^{3}P_{2}^{2}P_{3}^{0}\nonumber\\
&=&\epsilon_{0}\dfrac{|\vec{\textbf{q}}_{2}|}{2k.P_{1}}\lbrace 2k.P_{1}\sqrt{-a^{2}e^{2}+m_{l}^{2}+|\vec{\textbf{q}}_{2}|^{2}}+a^{2}e^{2}\omega -2n\omega k.P_{1} +2k.P_{1}\nonumber\\
&\times &\sqrt{|\vec{\textbf{q}}_{2}|^{2}-2|\vec{\textbf{q}}_{2}|\omega\cos\theta(n-\dfrac{e^{2}a^{2}}{2k.P_{1}})+\omega^{2}(n-\dfrac{e^{2}a^{2}}{2k.P_{1}})^{2}} \;\ \rbrace\sin\theta\sin\phi, \\
\varepsilon[a_{2},k,P_{2},P{3}]&=&\varepsilon_{2013}a_{2}^{2}k^{0}P_{2}^{1}P_{3}^{3}+\varepsilon_{2031}a_{2}^{2}k^{0}P_{2}^{3}P_{3}^{1}+\varepsilon_{2301}a_{2}^{2}k^{3}P_{2}^{0}P_{3}^{1}+\varepsilon_{2310}a_{2}^{2}k^{3}P_{2}^{1}P_{3}^{0}\nonumber\\
&=&-\epsilon_{0}\dfrac{|\vec{\textbf{q}}_{2}|}{2k.P_{1}}\lbrace 2k.P_{1}\sqrt{-a^{2}e^{2}+m_{l}^{2}+|\vec{\textbf{q}}_{2}|^{2}}+a^{2}e^{2}\omega -2n\omega k.P_{1} +2k.P_{1}\nonumber\\
&\times &\sqrt{|\vec{\textbf{q}}_{2}|^{2}-2|\vec{\textbf{q}}_{2}|\omega\cos\theta(n-\dfrac{e^{2}a^{2}}{2k.P_{1}})+\omega^{2}(n-\dfrac{e^{2}a^{2}}{2k.P_{1}})^{2}}\;\ \rbrace\sin\theta\cos\phi, \\
\varepsilon[a_{1},a_{2},k,P{1}]&=&\varepsilon_{1203}a_{1}^{1}a_{2}^{2}k^{3}P_{1}^{0}=\dfrac{\epsilon_{0}^{2}E_{1}}{\omega},\\
\varepsilon[a_{1},k,P{1},P_{2}]&=&\varepsilon_{1302}a_{1}^{1}k^{3}P_{1}^{0}P_{2}^{2}=\epsilon_{0}E_{1}|\vec{\textbf{q}}_{2}|\sin\theta\sin\phi, \\
\varepsilon[a_{2},k,P{1},P_{2}]&=&\varepsilon_{2301}a_{2}^{2}k^{3}P_{1}^{0}P_{2}^{1}=\epsilon_{0}E_{1}|\vec{\textbf{q}}_{2}|\sin\theta\cos\phi.
\end{eqnarray}

\end{document}